# Device-scaling constraints imposed by the van der Waals gap formed in two-dimensional materials

Mahdi Pourfath[1*] and Tibor Grasser[1*]

[1]Institute for Microelectronics, TU Wien, Gusshausstrasse 27-29/E360, 1040 Vienna, Austria.

[*]Corresponding author. Email: pourfath@iue.tuwien.ac.at and grasser@iue.tuwien.ac.at

**Transistor miniaturization requires controlling gate leakage through ultra-thin dielectrics and minimizing source/drain contact resistance. Although two-dimensional (2D) semiconductors offer excellent electrostatic control, their interfaces with gate dielectrics and contact metals often form a van der Waals (vdW) gap that impacts device performance and acts as a tunneling barrier with a low-dielectric constant. While this reduces dielectric leakage, it increases metal–channel contact resistance and introduces a parasitic series capacitance to the gate. We quantified the trade-off between leakage suppression and electrostatic and contact-resistance scaling limits. As a result, many insulators fail to meet scaling targets, and metal–channel contacts fall short of required resistances. Zipper-like interfaces, where quasi-covalent bonding removes the vdW gap without creating dangling bonds, offer a path toward ultrascaled transistor designs.**

Two-dimensional (2D) semiconductors, such as $MoS_2$ and other layered materials, are candidates for ultrascaled field-effect transistors (FETs) with channel lengths in the single-digit nanometer regime (5–10 nm) (*1*). Continued scaling to these dimensions also requires gate dielectrics with high permittivity (dielectric constant $\kappa > 15$) and large band offsets to suppress leakage relative to $SiO_2$ ($\kappa \sim 3.9$) (*2*). Gate dielectric performance is commonly quantified by the equivalent oxide thickness (EOT), which is the physical thickness of a $SiO_2$ dielectric layer with the same gate capacitance. According to the International Roadmap for Devices and Systems (IRDS) (*3*), beyond-2030 targets for ultrascaled transistors require a capacitance-equivalent thickness (CET) < 9 Å; accounting for the channel-side capacitance overhead, this translates to an insulator $\mathrm{EOT_{ins}}$ < 5 Å. Moreover, IRDS targets require a total source/drain resistance $R_{\mathrm{sd}} < 180\ \Omega\mu\mathrm{m}$. Whereas Si forms strong, fully covalent interfaces with $SiO_2$ and other dielectrics, 2D semiconductors form weaker interfaces with dielectric layers and source/drain contacts that are separated by a van der Waals (vdW) gap. We show that this vdW gap strongly affects both EOT and contact resistance.

Conventional screening through theoretical methods of candidate dielectrics typically focuses on bulk properties, such as permittivity and band gap, and assumes ideal, abrupt interfaces (*4, 5*), as illustrated in Fig. 1A. However, in real devices, several nonideal interfacial phenomena, such as reduced effective permittivity in thin films that can be caused by interface-induced dead layers (*6*), vdW gaps between the insulator and the channel, interface dipoles, and remote phonon scattering in high-$\kappa$ materials (*7*) (Fig. 1B), can substantially alter benchmarking conclusions.

One key limitation arises from the interfacial dead layers (*6*), commonly observed in high-$\kappa$ insulators. Structural or chemical perturbations, such as strain, defects, or incomplete crystallinity, can create thin regions with greatly reduced permittivity. These dead layers substantially increase the minimum achievable EOT for a given maximum leakage current by degrading the effective permittivity at small thicknesses (Fig. 1C).

Another critical factor is the weak vdW interaction (*8*) between 2D semiconductors and deposited insulators, which creates an interfacial vacuum-like gap that is usually overlooked. This gap produces a small but non-zero charge redistribution and exhibits weak polarization and an effective permittivity of about 2 to 3, which is meaningfully lower than that of the insulator. The vdW gap plays a dual role. It increases EOT by introducing a low-$\kappa$ series layer, but also suppresses tunneling by acting as an additional barrier. As shown in Fig. 1C, whether the vdW gap increases or reduces the minimum achievable EOT depends on which of these two effects dominates. For clarity, we focus specifically on the lower bound of the minimum achievable EOT at a fixed leakage target imposed by interfacial effects, most notably the vdW gap.

Although the vdW gap may seem negligible in size, we will show that even a sub-nanometer gap can substantially impact device performance. The tunneling current through a gate stack decreases exponentially with both the barrier height and the insulator thickness, scaled by its permittivity. In the Wentzel–Kramers–Brillouin (WKB) approximation, this exponential dependence is captured by the inverse decay length $\beta$, which depends on the tunneling effective mass and the relevant band offset $\Delta E$ between the insulator and the channel; $\beta \propto \sqrt{m^* \Delta E}$, with $m^*$ being the effective tunneling mass and $\Delta E$ the height of the tunneling barrier (details in the Supplementary Text, Section S4). To facilitate the screening and benchmarking of potential dielectrics, we define an insulator figure of merit (FoM) (*9*) as the ratio of $\kappa\beta$ for a candidate insulator relative to $SiO_2$ (Section S5):

$$\text{FoM} = \frac{\kappa_{\text{ins}} \beta_{\text{ins}}}{\kappa_{\text{SiO}_2} \beta_{\text{SiO}_2}}.$$

This metric quantifies how effectively an insulator can suppress tunneling while maintaining strong electrostatic control, relative to $SiO_2$. A higher FoM indicates lower leakage for the same EOT, enabling more aggressive scaling. The idealized case where a large permittivity is assumed to directly reduce EOT is shown in Fig. 1D. However, as shown below, the presence of a vdW gap can alter this simple model and in turn benchmarking outcomes.

An example showing that a high-$\kappa$ (high-FoM) insulator does not necessarily result in a low EOT is $SrTiO_3$ (STO). Figure 1B schematically shows an STO–$MoS_2$ heterostructure, where interfacial dead layers form near the boundaries. Their impact is captured by the degradation factor $D$ (an interfacial length parameter, defined in Eq. S22), that results in a thickness-dependent correction to the effective permittivity (*6*).Beyond the interfacial dead-layer capacitance, strong electric fields near interfaces can locally suppress the permittivity of high-$\kappa$ dielectrics such as STO (*10*), an effect captured by a field-dependent interfacial term (see Supplementary Text).

As the physical thickness of the insulator decreases, the relative impact of these regions becomes more pronounced. Although bulk STO exhibits a permittivity as high as $\kappa \approx 270$, ultrathin films fall short of ideal electrostatic performance because a large portion of the electric field drops across the low-$\kappa$ interfacial regions. Layer-resolved electronic structure and planar-averaged potential profiles (Fig. S1) reveal pronounced potential variations near the interface, indicating locally reduced permittivity. These interfacial regions therefore impose a fundamental penalty on the effective dielectric response of ultrathin gate stacks.

Fig. 1B summarizes these effects using an equivalent circuit model, where both the vdW gap and dead layers act as series capacitive bottlenecks that limit the minimum achievable EOT. As summarized in Table S2, for a wide range of insulator–$MoS_2$ stacks an average vdW gap contribution of $EOT_{vdW} \approx 2.7$ Å was obtained. Within the series-capacitance framework, this gap therefore amounts to ~30% of the 9 Å CET target projected by the IRDS for the 2030 technology node, even before accounting for the physical oxide thickness or the channel-side capacitance.

# Characteristics of the vdW gap

The vdW gap and its physical origin were first characterized using ab initio calculations, which revealed that the binding energy of a vdW–bonded interface between a 2D semiconductor channel and an insulator is typically one to two orders of magnitude smaller than that of a fully covalent interface (Fig. 2 A and B). For example, layered dielectrics such as hexagonal boron nitride (hBN) or perovskite oxide dielectrics like STO adhere to $MoS_2$ with typical surface binding energies of 15 to 30 meV/Å$^2$ (*11*), whereas a covalent Si–$SiO_2$ interface exhibits binding energies on the order of eV/Å$^2$. This weaker vdW bonding results in a larger equilibrium separation between materials (Fig. 2C). This separation creates a vacuum-like region between solids, where the crystal wavefunctions decay into evanescent tails.

However, not all interfaces are purely vdW or purely covalent. For example, the STO–$MoS_2$ interface appears more strongly vdW bonded than hBN–$MoS_2$, while certain structures such as $\beta$-$Bi_2SeO_5$–$Bi_2O_2Se$ ($\beta$-BSO–BOS) form zipper-like interfaces (*12*). These zipper-like bonds fall between the extremes of purely vdW and fully covalent bonding and are of particular interest because they may enable strong interfacial adhesion without introducing a vdW gap or creating dangling bonds which, as will be shown bellow, is an essential characteristic for scalable device integration. A comparison of the vdW-bonded (Fig. 2A) and covalent interfaces (Fig. 2C), shows both the lower binding energy and the more uniform interfacial charge in vdW systems.

We define the vdW gap, $t_{vdW}$, as the distance between adjacent atomic planes minus the sum of their covalent radii (Fig. 2B), thereby isolating the vacuum-like region between the two solids rather than the full core-to-core spacing. The equilibrium positions and relative distances between neighboring atoms in a vdW stack can be obtained from *ab initio* calculations and can also be inferred from high-resolution TEM imaging. The vdW gap is then extracted by subtracting the corresponding covalent radii from these interatomic distances.

Across the insulator–$MoS_2$ interfaces examined in this work, the equilibrium vdW gap thickness is narrowly distributed, with a characteristic value of $\bar{t}_{\mathrm{vdW}} \approx 1.4$ Å (Table Table S2). To assess the generality of this scale beyond the specific systems studied here, we performed an independent estimate of vdW gaps based on tabulated atomic radii (Materials and Methods). This analysis yielded a statistical distribution of vdW gap thicknesses in the range of 1 to 2 Å, with a mean value of $1.40 \pm 0.22$ Å, as shown in Fig. 2D.

## Dielectric properties of vdW gaps

As already implied, the vdW gap acts as a low-permittivity interfacial layer that substantially reduces the out-of-plane dielectric response of 2D heterostructures. The same framework can be extended to layered 2D dielectrics (see Supplementary Text), where internal vdW gaps exist between atomic layers (*13*). To describe spatial variations in dielectric screening across the heterostructure, we use an effective permittivity profile, $\kappa(z)$, which provides a compact representation of local polarization response. Device-relevant capacitance and EOT are obtained from this profile through appropriate spatial averaging (Materials and Methods).

Fig. 3, A to C, focuses on an hBN–$MoS_2$ heterostructure, a widely studied 2D gate stack (*14*), and illustrates how the vdW gap reduces the effective permittivity. For reference, the responses of isolated hBN and isolated monolayer $MoS_2$ are plotted as well. An enhancement of polarization in the gap region was observed (Fig. 3B) in the heterostructure compared to the isolated cases, corresponding to a non-uniform $\kappa(z)$ that attained a minimum within the vdW gap (Fig. 3C). Physically, the gap behaves nearly like vacuum, with only modest polarizability from evanescent electronic tails near the interfaces (Materials and Methods). For the hBN–$MoS_2$ stack, the effective vdW-gap permittivity $\kappa_{\mathrm{vdW}}^{\mathrm{eff}}$ was $\approx 1.7$. Across a range of representative systems, the average value of $\bar{\kappa}_{\mathrm{vdW}}$ was $\approx 2$ (Table S2). Such low-$\kappa$ interfacial layers weaken electrostatic coupling and systematically increase the EOT, imposing a fundamental constraint on dielectric scaling even in the presence of high-$\kappa$ materials.

A practical consequence of the vdW gap's low permittivity is that it introduces a nearly constant series thickness to the total EOT, regardless of the bulk insulator. Across a range of insulator–$MoS_2$ interfaces analyzed using DFT-relaxed geometries, the equilibrium vdW gap thickness, $t_{\mathrm{vdW}}$, varied moderately (typically within $\pm 0.2$ Å), and the extracted permittivity exhibited a clear dependence on $t_{\mathrm{vdW}}$ (Fig. 3D and E; see also Fig. S3). This behavior is well described by a simple analytical model relating the effective vdW-gap permittivity to the interfacial separation (Materials and Methods).

The model provides a reliable estimate of how variations in interface spacing (such as those induced by fabrication-related strain or surface roughness) may affect $\kappa_{\mathrm{vdW}}$. On average, the vdW gap contributes an effective permittivity of $\kappa_{\mathrm{vdW}} \sim 2$ and an equivalent oxide thickness of $\mathrm{EOT}_{\mathrm{vdW}} \approx 2.7$ Å, as shown in Fig. 3 and listed in Table S2. These results indicate that more than a quarter nanometer of EOT is intrinsically lost to the vdW gap in typical 2D transistors, a substantial penalty when striving for subnanometer scaling in light of the IRDS roadmap (*3*) targets discussed below.

By analogy to Si CMOS practice, even in high-$\kappa$ $HfO_2$ stacks, a thin interfacial $SiO_2$ layer is retained to preserve interface quality and mobility, which also imposes a non-negligible EOT penalty, and has been an established trade-off in high-$\kappa$/metal–gate integration (*7*). In vdW heterostructures, the vdW gap plays the analogous interfacial role; its impact on tunneling is discussed next.

Although vdW gaps appear at 2D channel–insulator interfaces, they also exist within layered materials themselves, where internal gaps between atomic planes reduce the bulk permittivity. Figure 3C illustrates that in multilayer hBN and $MoS_2$, the permittivity profile $\kappa(z)$ peaks near atomic layers and drops in the vdW gaps between layers. A simple analytical model treats a layered crystal as a sequence of alternating high-$\kappa$ slabs (representing atomic planes) and low-$\kappa$ vdW gaps.

For example, an $N$-layer $MoS_2$ stack can be modeled as two surface half-layers plus $(N-1)$ internal layer–gap bilayers arranged in series (see Materials and Methods), which reproduces the increase of the out-of-plane permittivity with $N$ and its convergence toward the bulk limit as $N \to \infty$ (Fig. 3F). A reduction in the interlayer gap, for example from strain or pressure, is therefore predicted to enhance the effective permittivity. Interestingly, one experimental study (*15*), also shown in Fig. 3F, reported an anomalously large out-of-plane permittivity in few-layer $MoS_2$, possibly caused by unintentional strain that narrows the vdW gaps.

# Tunneling through vdW gaps

The vdW gap is essentially a nanoscale vacuum barrier that has a profound impact on electron tunneling between the channel and the gate. Both analytical WKB estimates within the Tsu–Esaki tunneling model and explicit quantum transport simulations illustrate this effect (Fig. 4 A and B). A representative graphene–hBN–graphene stack serves as a model system to benchmark the tunneling model against experimental conditions. In this structure, the planar-averaged electrostatic potential across the vdW gap approaches the vacuum level (Fig. 4B), resulting in an effective barrier height for electrons tunneling into graphene that is approximately equal to its work function (~4.5 eV (*16*)).

We applied the WKB approximation to analytically estimate the tunneling probability and capture the exponential suppression of transmission with increasing vdW gap thickness. For thin insulating barriers, coherent transport can be assumed, and the total transmission can be approximated as the product of the contributions from the bulk insulator and the vdW gap. To account for deviations from an ideal rectangular tunneling barrier under realistic electrostatics, we introduce a dimensionless barrier-shape factor $\alpha$. Whereas $\alpha = 1$ corresponds to a rectangular profile, smoother, more trapezoidal barriers can reduce $\alpha$ toward $2/3$. In practice, the appropriate value of $\alpha$ depends on the vdW gap geometry, with larger gaps approaching the rectangular limit and narrower or perturbed gaps exhibiting softer barriers (Materials and Methods).

For the graphene–hBN–graphene system, the DFT-relaxed vdW gap is $\sim$1.6 Å, yielding a single-gap transmission probability of $T_{\mathrm{vdW}} \approx 6.2\%$. Because there are two vdW gaps on the opposing sides of the hBN layer, the overall transmission is approximately the product of the two, leading to an estimated $\sim$ 260-fold suppression of the tunneling current. This substantial reduction occurs even though the vdW gap is atomically thin, because its barrier height is nearly as high as the vacuum level. Figure 4A confirms this effect: ab initio non-equilibrium Green's function (NEGF) quantum transport simulations show that the local density of states (LDOS) decays sharply within hBN and drops substantially at the graphene-hBN interfaces, particularly near the graphene Fermi level. Overall, the calculated transmission remains consistent with the WKB estimate (Materials and Methods; Section S7).

The proposed model can be compared to experimental data (*17*) on tunneling through 2D insulators. Figure 4 C shows the current-voltage characteristics for graphene-hBN-graphene structures with one to four layers of hBN as the tunneling barrier. The key simulation parameters are described in detail in the Supplementary Materials. The solid lines (model) and symbols (experiment) show excellent agreement across all thicknesses when the effect of the vdW gaps is included. Without accounting for the gaps, the measured resistance versus hBN thickness is underpredicted by orders of magnitude. Including the two vdW gaps in the simulation increases the calculated resistance by nearly 260×, precisely matching the experimental trend [Fig. 4D]. A slight deviation remains for monolayer hBN at high bias, but this is resolved by including a small series resistance in the model (red dashed line in Fig. 4C), which accounts for voltage drops in the contacts at high current levels.

Figure 4E compares the developed model with experimental data for a gate stack composed of a monolayer $MoS_2$ channel, an STO insulator, and an Au metal gate (*18*). Because $MoS_2$ is the channel material in most 2D transistors considered here, its electron affinity ~4.3 eV determines the effective tunneling barrier into the conduction band. Applying a simple WKB model across a vdW gap of ∼1 to 2 Å, this barrier yielded a transmission probability of only ~3 to 20%, depending on the gap width and the barrier shape. A single vdW gap at the STO-$MoS_2$ interface, estimated in this work to be ∼1.4 Å, resulted in a transmission probability of $T_{\mathrm{vdW}} \approx 9.2\%$. Although the details of the Au-STO interface were not provided in the experimental data, the assumption a transmission probability of 20% (representative of the smallest vdW gaps) yielded excellent agreement with experiment (*18*).

Another relevant scenario is the metal–2D semiconductor contact, which plays a critical role in enabling low source and drain contact resistances to the channel. In the absence of any vdW gap, such contact s can approach the quantum resistance limit, given by $R_{\mathrm{cq}}W = \frac{h}{2e^2}\sqrt{\frac{\pi}{2}}\,\frac{1}{\sqrt{g_v\, n_{2D}}}$ (*19*), where the valley degeneracy $g_v = 2$ for monolayer $MoS_2$ is the result of its two equivalent valleys in the Brillouin zone. This expression assumes unit interfacial transmission ($T = 1$). Here we generalize it to realistic metal–2D interfaces by explicitly including the vdW gap as a vacuum-like tunneling barrier with transmission $T_{\mathrm{vdW}} < 1$ [see Supplementary Text, Section S6, Eq. S74].

When a metal electrode is placed on a 2D semiconductor, a vdW gap of a few angstroms typically exists, unless specific processes (such as surface treatments or interfacial alloying) promote covalent bondin g and eliminate the gap. Using the generalized form with $T_{\mathrm{vdW}} < 1$, we evaluate the quantum contact resistance for metal–$MoS_2$ across vdW gap thicknesses and barrier-shape factors ($\alpha \in [2/3, 1]$) (Fig. 4F; quantitative details in Section S6). The vdW gap range considered here slightly exceeded the typical equilibrium values discussed above, because in practice, many device fabrication methods (such as transfer printing of contacts) can inadvertently introduce larger-than-equilibrium gaps at interfaces, making this issue particularly relevant. For wider gaps, the barrier profile becomes more rectangular, making $\alpha \to 1$ appropriate. Even modest increases in $t_{\mathrm{vdW}}$ produce orders-of-magnitude increases in $R_c W$, and the model aligned with experimental reports for purely vdW-bonded metal–$MoS_2$ contacts (*20–26*)

Figure 4F places these results in the context of IRDS targets for source–drain resistance. For technology nodes beyond 2030, the IRDS roadmap requires the total source and drain contact resistance $R_{SD}$ to be reduced to 180 Ωμm, with the latter target indicated by the dashed horizontal line. As shown, this requirement cannot be met in the presence of a finite vdW gap under realistic conditions. Only in the limiting case of high degeneracy in the 2D semiconductor, combined with a reduced or eliminated vdW gap, does $R_{cq} W$ approach the IRDS target.

## Scaling limits imposed by vdW gaps

To place 2D transistor scaling in the context of established Si technology, it is instructive to partition the total gate-stack CET into channel and insulator contributions. In bulk Si MOSFETs, the inversion-layer centroid displacement and quantum-confinement effects introduce an effective electrostatic thickness $\sim 4$ Å, such that the IRDS beyond-2030 target of $CET_{total} < 9$ Å leaves only 5 Å for the gate insulator EOT. In Si, this channel contribution is well established from prior theory and experiment.

For atomically thin 2D semiconductors, the corresponding channel contribution must be re-evaluated explicitly, particularly in the presence of finite quantum capacitance and a vdW gap. As shown in Section S8, we extracted the $MoS_2$ channel capacitance self-consistently from DFT-derived local density of states and charge-centroid profiles, yielding a representative channel contribution of $CET_{MoS_2} \approx 3$ Å at $V_G - V_{th} \approx 0.4$ V, consistent with IRDS projected operating conditions. Consequently, meeting the IRDS target requires the combined insulator and vdW gap EOT to remain $< 6$ Å.

Integration of the electrostatic and tunneling effects of vdW gaps enables evaluation of their impact on the scaling limits of various gate insulators. These competing effects are shown in Fig. 5A, which summarizes leakage current versus EOT curves for two example dielectrics, hBN and STO, with $MoS_2$ as the channel. The calculated $J$–EOT relation for hBN–$MoS_2$ is presented under three different scenarios: (i) ignoring the vdW gap entirely (only the bulk insulator considered); (ii) including the tunneling barrier effect of the vdW gap but not its capacitance penalty; and (iii) including both the barrier and the permittivity cost of the vdW gap (realistic model).

In the case of hBN (modest-$\kappa$, wide-band-gap) (*27*), the dotted curve (tunneling only) shows that the additional barrier reduces leakage compared to the dashed curve (no gap). This shift enables scaling to a smaller EOT for a given $J_{\mathrm{target}}$ (target leakage current density). However, when the finite permittivity of the gap is included (solid curve), the EOT cannot actually reach competitive values. In the case of hBN, the net outcome remains positive. The lowest attainable EOT with the gap included is slightly smaller than it would be without the gap, meaning the leakage suppression outweighed the capacitance drawback. We remark that this slight improvement is specific to hBN's modest $\kappa$ (low FoM); for all other cases studied, the vdW gap worsens the minimum achievable EOT.

For STO (ultra-high $\kappa$ and moderate band offset), the situation is reversed. If interface effects are erroneously neglected, STO appears capable of achieving extremely small EOT values (well below 6.5 Å) before reaching the leakage floor (gray curve). However, in practice, STO films exhibit appreciable dead layer effects that increase the total minimum achievable EOT. In the limiting case where the electronic contribution to the permittivity is negligible, Eq. S53 implies that the dead layer offset to the minimum EOT is approximately $D\kappa_{\mathrm{SiO_2}}$.

Using a dead-layer parameter of $D = 1.61$ Å (*18*), which is consistent with thin-film series-capacitance analyses that infer an interfacial thickness scale of $\sim$ 1–2 Å (*28*), the minimum achievable EOT is increased to approximately 5 Å. For STO with a physical thickness of $\sim$ 45 Å (10 unit cells), this model yields an effective permittivity of $\sim$ 30, corresponding to EOT $\approx$ 5.8 Å (*18*). At an operating bias of 0.6 V (IRDS requirement), a portion of $V_{\mathrm{G}}$ reduces across the vdW gap because of its lower permittivity, so the electric field inside STO is modest ($E_i \approx$ 0.3 to 0.4 MV/cm), a regime where bulk field-dependent data still indicate larger permittivities (*10*). Hence, the interfacial series term dominates; however, in general, field-induced permittivity reduction can further increase $\mathrm{EOT}_{\mathrm{min}}$.

In addition, the vdW gap adds an extra series thickness. When these effects are included in the realistic model, the minimum achievable EOT is much larger. Indeed, under the parameters used in this study (see Supplementary Materials), the STO stack failed to reach the IRDS target of ~6 Å EOT. Thus, the interfacial dead layer and vdW gap worsen the scaling limit, forfeiting the advantage STO would otherwise offer because of its very high bulk $\kappa$. However, the zippered $\beta$-BSO–BOS stack forms continuous interfacial bonding with the channel (neither purely vdW-bonded nor fully covalent), shows no such penalty, and can, in principle, meet the sub-5 Å EOT target, as verified experimentally by achieving a minimum EOT of 4 Å (*29*). These examples demonstrate that vdW interface effects can alter a material's suitability for extreme scaling.

Figure 5B compiles the minimum achievable EOT for a range of gate dielectrics, comparing the values obtained with and without considering the vdW gap. The vdW gap values used in this study are the ideal equilibrium values obtained from ab initio calculations. However, in practice, the vdW gap can be larger because of interface roughness and imperfections (*30*) or fabrication-related effects (*31*), with values in the range of 0.6 to 2 Å or more being possible. Therefore, to analyze more realistic scenarios, this study considers both the ab initio–calculated vdW gap and a case where the gap is increased by 50%. The enlarged-gap case includes a thickness-dependent reduction of the vdW gap permittivity (parameters in Table S2), which amplifies the series EOT penalty: $t_{\mathrm{vdW}} = 1.42$ Å gives $\mathrm{EOT}_{\mathrm{vdW}} \approx 2.7$ Å, whereas a 50% increase in the vdW gap thickness yields $\mathrm{EOT}_{\mathrm{vdW}} \approx 5.4$ Å, consistent with reduced interfacial electronic polarizability at larger separations.

Given the variability in reported values of electron affinity, tunneling effective mass, and permittivity, each material is represented by a range of minimum achievable EOT values. For each case, reported values and physically reasonable parameters (see Supplementary Materials) were used to compute the tunneling current as a function of EOT; the procedure is summarized in Fig. S4. The intersection of these curves with the target leakage current defines the corresponding range of minimum achievable EOTs. In nearly all cases, inclusion of a vdW gap shifts the minimum achievable EOT to higher values. For high-$\kappa$ insulators, this shift can critically limit their viability, pushing the EOT beyond the required acceptable limits. For instance, $CaF_2$ and STO both end up above 6 Å EOT when the vdW gap is included, whereas without the gap they may seem to be able to reach 6 Å.

A key limitation of this analysis is that thickness-dependent permittivity data, $\kappa(t_{\mathrm{ins}})$, are largely unavailable for most candidate insulators. With the notable exceptions of $HfO_2$ (*32*) and STO (*18*), experimental information on the reduction of dielectric response in the ultrathin limit is scarce. Consequently, for the remaining materials the projections implicitly assume a best-case scenario corresponding to an ideal dielectric with no additional thickness-induced degradation ($D = 0$), and the resulting minimum achievable EOT values should be interpreted as optimistic lower bounds.

For $HfO_2$, the reported thickness-dependent permittivity $\kappa(t_{ins})$ was extracted from metal–$HfO_2$–$MoS_2$ metal–insulator–semiconductor (MIS) gate stacks (*32*), in which the presence of a vdW gap at the $HfO_2$–$MoS_2$ interface is plausible. We therefore interpret the measured electrical thickness as comprising both the oxide layer and an interfacial vdW contribution, and correct the reported permittivity using a series-capacitance formulation (Fig. S5C and Eq. S23). By contrast, the STO data were obtained from metal–insulator–metal (MIM) capacitors (*18*), where conformal metal–oxide bonding makes the existence of an interfacial vdW gap unlikely. In this case, the observed thickness dependence is attributed to intrinsic dead-layer effects, and no vdW correction is applied. If future evidence were to indicate the presence of interfacial gaps, the same correction procedure could be straightforwardly extended to STO or other materials.

Within our modified dead-layer model, only the ionic contribution to the dielectric response is assumed to degrade with decreasing thickness, while the electronic polarizability remains unaffected. This reflects the greater sensitivity of ionic softness to confinement and interfacial perturbations compared with the electronic background (Eq. S55).

An important exception to this general trend occurs for modest-$\kappa$ materials such as hBN, where the suppressed leakage caused by the vdW gap can result in a somewhat smaller final EOT than would occur otherwise. However, because of the intrinsically low-$\kappa$ of such insulators, they remain promising candidates primarily in applications where extreme leakage suppression outweighs the capacitance penalty. A possible compromise appears to be $LaF_3$, which offers an intermediate permittivity of $\sim$13 (*33*) and favorable band alignment. Although the inclusion of the vdW gap increases its minimum EOT from $\sim$1.7 Å to 3.9 Å, a notable penalty, the resulting values remain highly attractive and could still meet projected requirements. In the presence of a dead layer for this material, $LaF_3$ may no longer satisfy the minimum EOT condition. However, as the electronic contribution to the permittivity increases, the relative impact of interfacial dead layer degradation diminishes, because interface effects predominantly suppress the ionic response and leave the electronic part largely unaffected. As a result, materials with a higher fraction of electronic polarizability are more resilient to interface-induced permittivity loss and can potentially support further EOT scaling.

These trends can be understood quantitatively by examining the minimum achievable EOT as a function of the insulator FoM. Using $MoS_2$ as a representative 2D channel material, we extract a vdW gap figure of merit $\text{FoM}_{\text{vdW}} \approx 1$ (Eq. S69), comparable to that of $SiO_2$. This reinforces the analogy with traditional Si CMOS, where an interfacial $SiO_2$ layer imposes an unavoidable EOT penalty (*34*). For realistic device stacks, the vdW gap imposes an approximately material-independent EOT offset, and any additional dielectric dead-layer effect further increases the minimum achievable EOT. These contributions can be captured in a compact analytical expression (Eq. S66, Eq. S69, and Eq. S53):

$$\text{EOT}_{\text{total}}^{\text{min}} \approx \frac{7\ \text{Å}}{\text{FoM}_{\text{ins}}} + 2.7\ \text{Å} + 3.9D, \tag{1}$$

where $D$ is the dead-layer parameter (in Å) accounting for thickness-dependent permittivity degradation. This analysis underscores the stringent FoM demands for future insulators and highlights the critical impact of vdW and dead-layer contributions. In the idealized case where dead-layer degradation is absent ($D = 0$), the condition EOTtotal = EOTins + EOTvdW < 6 Å is satisfied when FoMins $\gtrsim$ 2. However, even a modest degradation of $D = 0.5$ Å tightens this requirement, raising the FoM threshold to $\text{FoM}_{\text{ins}} \gtrsim 5$. As shown in Fig. 5C, relatively few candidate oxides exceed this higher threshold, illustrating how the combined effect of the vdW interface and dead-layer penalty imposes severe constraints on scalable gate stack design.

In the earlier examples, hBN (with mediocre $\kappa_{\text{ins}}$ and modest band offset) yielded a $\text{FoM}_{\text{ins}}$ between 0.8 and 1.2, suggesting that the vdW gap can be beneficial. For $LaF_3$, $\text{FoM}_{\text{ins}}$ was between 5 and 7, far above the vdW gap's FoM, so the gap eroded its scaling advantage. Figure 5C shows the calculated minimum EOT as a function of the insulator FoM, both with and without a vdW gap. Both the analytical model (Eq. S70) and numerical calculations based on the Tsu–Esaki model confirmed that the vdW gap imposes its largest EOT penalty for high–FoM insulators. In such cases, the minimum achievable EOT asymptotically approaches the vdW gap-only limit.

Figure 5D quantifies the electrostatic budget imposed by the $MoS_2$ channel contribution to CET. For the IRDS target $\text{CET}_{\text{total}} < 9$ Å, the extracted $\text{CET}_{MoS_2} \approx 3$ Å at $V_G - V_{\text{th}} \approx 0.4$ V leaves < 6 Å for the combined insulator and vdW gap EOT. For a DFT-predicted vdW gap, this implies an insulator EOT budget of < 3 Å, whereas realistic enlarged gaps leave < 1 Å for the insulator, an impractically stringent requirement. Thus, even moderate deviations from ideal vdW gap control can completely exhaust the EOT budget, establishing the vdW gap as the dominant scaling bottleneck for 2D gate stacks.

# Implications and outlook

The vdW gap can either improve or impede scaling of 2D semiconductor devices. It can provide a substantial advantage by suppressing direct tunneling across gate insulators, where it acts as a vacuum-like barrier that can reduce leakage currents by up to two orders of magnitude. However, the vdW gap can introduce a severe electrostatic penalty. Even a sub-nanometer vdW gap with its modest permittivity can add an EOT of ~2.7 Å, undermining the benefits of high-permittivity materials and limiting overall gate control. We note that the EOT penalty introduced by the vdW gap is both analogous to and comparable in magnitude to the EOT penalty introduced by the thin $SiO_2$ layer in Si technology.

While monolayer 2D channels like $MoS_2$ indeed offer inherently strong electrostatics that can relax short-channel constraints, gate dielectric scaling remains essential for high-performance 2D logic. From a circuit-level perspective, total gate capacitance must be maximized to achieve the required on-state carrier density and drive current at low operating voltages. Under demanding scaling targets, the vdW gap at the 2D channel/dielectric interface becomes a limiting factor, acting as an additional dielectric layer that adds ~2.7 Å to the EOT. In other words, a noticeable fraction of the allowable EOT is consumed by the non-bonded interface alone, leaving less room for high-$\kappa$ insulators. Moreover, the substantial increase in source/drain resistance due to the vdW gap at contact interfaces further limits the ability of 2D FETs to meet IRDS performance targets. Going forward, continued innovations in interface engineering—to minimize or eliminate the vdW gap—will be essential for enabling 2D materials to fully leverage their electrostatic advantages under realistic scaling constraints.

These results challenge the prevailing assumption that a high bulk permittivity and large band gap are sufficient for an insulator to support deep scaling. In practice, the presence of interfacial effects such as vdW gaps and dead layers can dominate the effective electrostatics and must be explicitly considered. These effects set a lower bound on the achievable EOT, regardless of how favorable the bulk dielectric properties may be. In this context, beyond the parallel-plate limit, fringing fields become a dominant non-ideality in ultrascaled devices, particularly penalizing thick high-$\kappa$ stacks through fringing-induced barrier lowering (FIBL). Although a low-$\kappa$ vdW gap can reduce the out-of-plane capacitance and increase the EOT, as discussed above, it can beneficially redirect lateral electric-field lines and thereby mitigate FIBL in short-channel geometries (*18, 35*). This further highlights the dual role of vdW gaps in the scaling landscape: they add vertical EOT yet can improve in-plane electrostatics when fringing dominates.

Taken together, these results imply that ultrahigh-$\kappa$ dielectrics whose permittivity is dominated by lattice (ionic) polarization are intrinsically vulnerable to interface- and field-induced suppression at nanometer thicknesses, whereas materials with a larger electronic polarization fraction are comparatively robust; the latter should therefore be prioritized for aggressive EOT targets.

To overcome the vdW gap bottleneck, new interfacial engineering strategies are likely needed. One potential strategy involves the application of strain to the insulator. The out-of-plane permittivity of vdW insulators is largely limited by the vdW gap between adjacent layers, whereas the permittivity in the vicinity of the atomic layers is typically much higher. The overall permittivity can be enhanced by reducing the interlayer vdW gap, for example, through the application of out-of-plane strain. This approach may offer a viable pathway for permittivity enhancement in layered insulators.

Another particularly promising approach is the use of zipper-like interfaces—engineered to transition from purely vdW to partially covalent bonding. This interfacial bonding continuity eliminates the vacuum-like gap and strengthens dielectric coupling. Native oxide approaches, where crystalline insulators are lattice-matched to the 2D semiconductor, exemplify this strategy. For example, the $\beta$-BSO–BOS system demonstrates $<5$ Å EOT without interfacial gaps (*29*), offering a proof-of-concept for scalable zipper material integration. A schematic illustration of this concept is shown in Fig. S6.

Eliminating the vdW gap (such as with zipper-like interfaces) is a powerful lever for EOT scaling, but device viability also hinges on co-optimizing channel transport and contacts. Beyond gate electrostatics, the vdW gap fundamentally limits contact resistance and, under realistic conditions, precludes meeting IRDS source–drain resistance targets, making its elimination essential for low-resistance 2D devices. Notably, BSO–BOS has demonstrated competitive room-temperature mobility (up to ~270 $\mathrm{cm^2\,V^{-1}\,s^{-1}}$) in vertical 2D FinFET-like devices, together with robust electrical characteristics (*36*); moreover, BOS fully encapsulated by BSO reaches $\sim$ 812 $\mathrm{cm^2\,V^{-1}\,s^{-1}}$ Hall mobility at 300 K (*37*), indicating that zipper-like integration is compatible with high room-temperature mobility.

The framework developed in this work provides a quantitative basis for incorporating such interfacial phenomena directly into design strategies. By extending the insulator FoM to include interface-specific factors, such as vdW gaps and interfacial permittivity suppression, more realistic scaling predictions can be made. Materials once considered ideal may prove unsuitable under these refined criteria, while others, previously overlooked, may emerge as viable candidates for next-generation devices.

Technology scaling toward gate-all-around nanosheet and nanowire architectures implies a transition from nominally two-dimensional to quasi-one-dimensional channel electrostatics. In this context, the present conclusions regarding vdW-gap and dead-layer–induced series capacitance and their impact on the minimum achievable EOT remain valid, whereas the absolute gate-leakage magnitude becomes geometry-dependent and is influenced by channel perimeter and curvature in gate-all-around devices.

The present analysis focuses on an idealized limit of high-quality interfaces, capturing intrinsic electrostatic and tunneling constraints arising from vdW gaps and dead layers. In practical devices, additional non-idealities, such as point defects, chemical disorder, and unintentional impurities in both the channel and the dielectric, can introduce defect-mediated transport mechanisms such as trap-assisted tunneling, which may substantially increase gate leakage beyond the direct-tunneling limits identified here. Thickness non-uniformity in the vdW gap and dielectric can further exacerbate leakage, since tunneling currents depend exponentially on local thickness and electric field, even when the corresponding variation in effective oxide thickness is modest. Accordingly, the $\mathrm{EOT}_{\mathrm{min}}$ values extracted in this work should be interpreted as intrinsic, best-case scaling limits, providing a baseline against which defect- and variability-limited device performance can be assessed in future studies.

Ultimately, overcoming the limitations imposed by vdW gaps is essential to realizing the full potential of 2D semiconductors in nanoscale electronics. By eliminating the interfacial vacuum gap and restoring dielectric continuity, zipper-like interfaces offer a promising pathway to approach the scaling limits of electrostatic control in 2D semiconductor devices.

Materials and Methods are available in the Supplementary Materials.

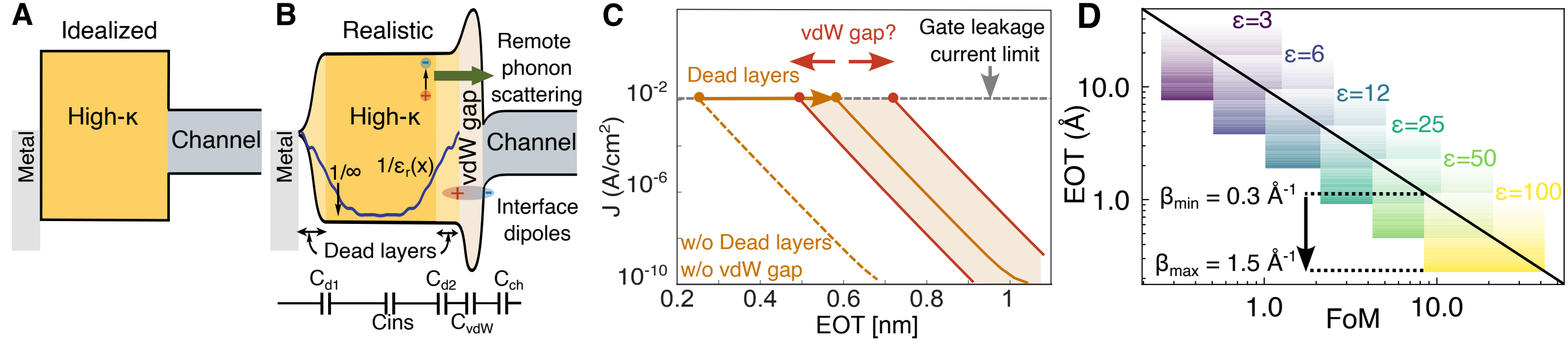

**Figure 1**: **Interface effects in gate–insulator–channel stacks.** **(A)** Idealized gate–insulator–channel stack without interfacial effects. **(B)** Realistic stack highlighting vdW gaps, interfacial dead layers, interface dipoles, and remote phonon scattering. The accompanying schematic circuit model illustrates how dead layers and vdW gaps act as series capacitors, increasing the total EOT. **(C)** Gate tunneling current density as a function of EOT. The minimum achievable EOT is determined by the intersection of the tunneling current with the maximum allowable leakage for a given application. Interfacial dead layers increase this minimum EOT by reducing the effective permittivity of ultrathin films, while the vdW gap simultaneously suppresses tunneling and adds electrostatic thickness, with either effect potentially dominating depending on material properties. **(D)** Minimum achievable EOT as a function of the insulator figure of merit (FoM), based on the analytical model described in Section S5. Higher FoM values enable scaling to subnanometer EOT. Because the FoM depends on both the dielectric permittivity and the inverse decay length ($\beta$), a representative range of $\beta$ values from 0.3 $Å^{-1}$ (leaky insulators) to 1.5 $Å^{-1}$ (good insulators) is shown.

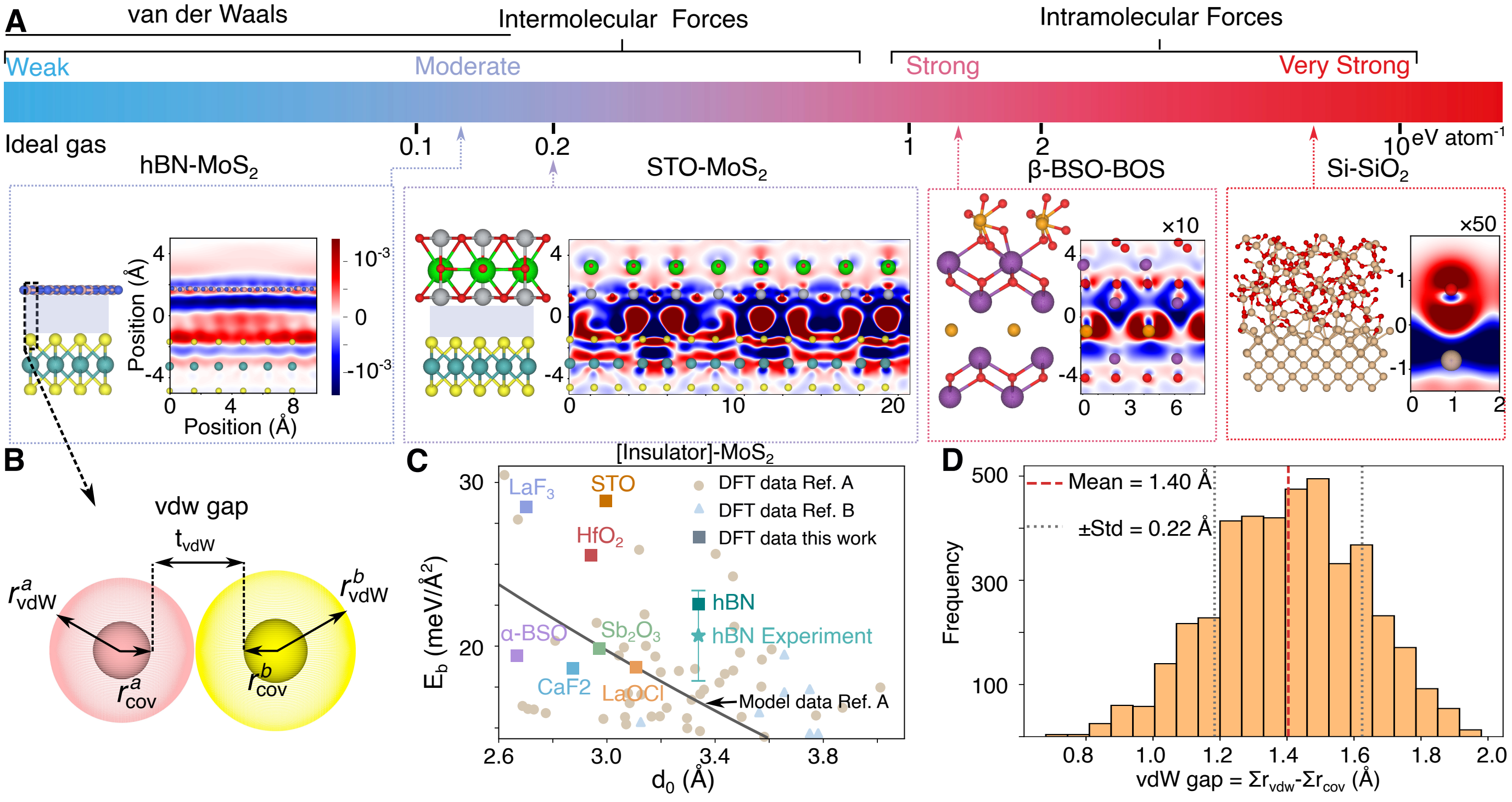

**Figure 2**: **vdW gap definition and statistics. (A)** Schematic comparison of interfacial bonding in vdW- and covalently bonded systems. vdW-bonded interfaces (such as hBN–$MoS_2$) exhibit weak binding and delocalized charge density, whereas covalent interfaces (such as $SiO_2$–Si) show strong binding with localized, directional charge. STO–$MoS_2$ exhibits intermediate bonding, while $\beta$-BSO–BOS forms a partially bonded (zippered) interface. **(B)** Definition of the vdW gap as the separation between two surfaces minus the sum of their atomic (covalent) radii, representing a vacuum-like interfacial region. **(C)** Binding energy versus vdW gap distance for representative 2D heterostructures. Smaller gaps generally correlate with stronger binding, with deviations arising from ionic character or enhanced vdW overlap. DFT data are sourced from Reference A (*38*) and Reference B (*39*). **(D)** Statistical distribution of vdW gap reference lengths derived from tabulated covalent and vdW radii (Table S1 and Fig. S2), yielding a mean value of 1.40 ± 0.22 Å.

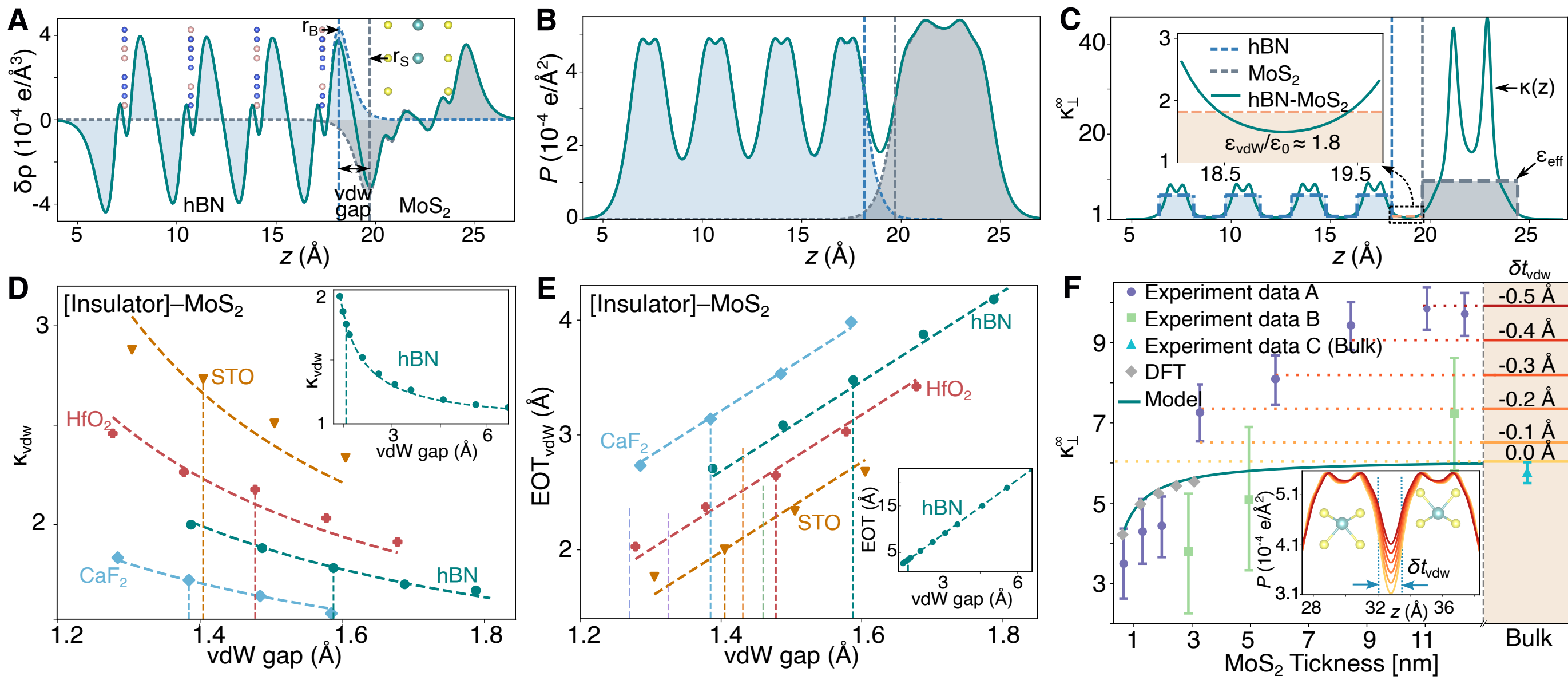

**Figure 3**: **Dielectric properties of the vdW gap.** **(A)** Induced charge density $\Delta\rho(z)$ in an hBN–$MoS_2$ heterostructure under a small out-of-plane electric field. Isolated $MoS_2$ (gray) and hBN (blue) are compared with the combined stack (teal), showing interfacial charge redistribution within the vdW gap (vertical dashed lines). **(B)** Induced polarization $P(z)$ obtained from $\Delta\rho(z)$, revealing finite polarization within the vdW gap in the vdW-bonded heterostructure. **(C)** Spatial permittivity profile $\kappa(z)$, which peaks near atomic planes and decreases toward unity within the vdW gap. The profile is modeled as alternating high-permittivity atomic layers and low-permittivity vdW regions. **(D)** Effective vdW-gap permittivity versus gap thickness for selected insulator–$MoS_2$ stacks from first-principles calculations. Narrower gaps exhibit enhanced polarization and higher effective permittivity. Vertical dashed lines mark equilibrium vdW gap thicknesses; inset shows hBN at enlarged separations. **(E)** EOT contribution of the vdW gap for the same systems, increasing approximately linearly with gap thickness. An equilibrium gap of ~1.4 Å contributes ~2.7 Å to the total EOT. **(F)** Effective out-of-plane permittivity of multilayer $MoS_2$ as a function of layer number, approaching the bulk limit. Reducing the interlayer vdW gap enhances polarization and increases the effective permittivity (inset). Experimental data are from (*15*) (A), (*40*) (B), and (*41*) (C).

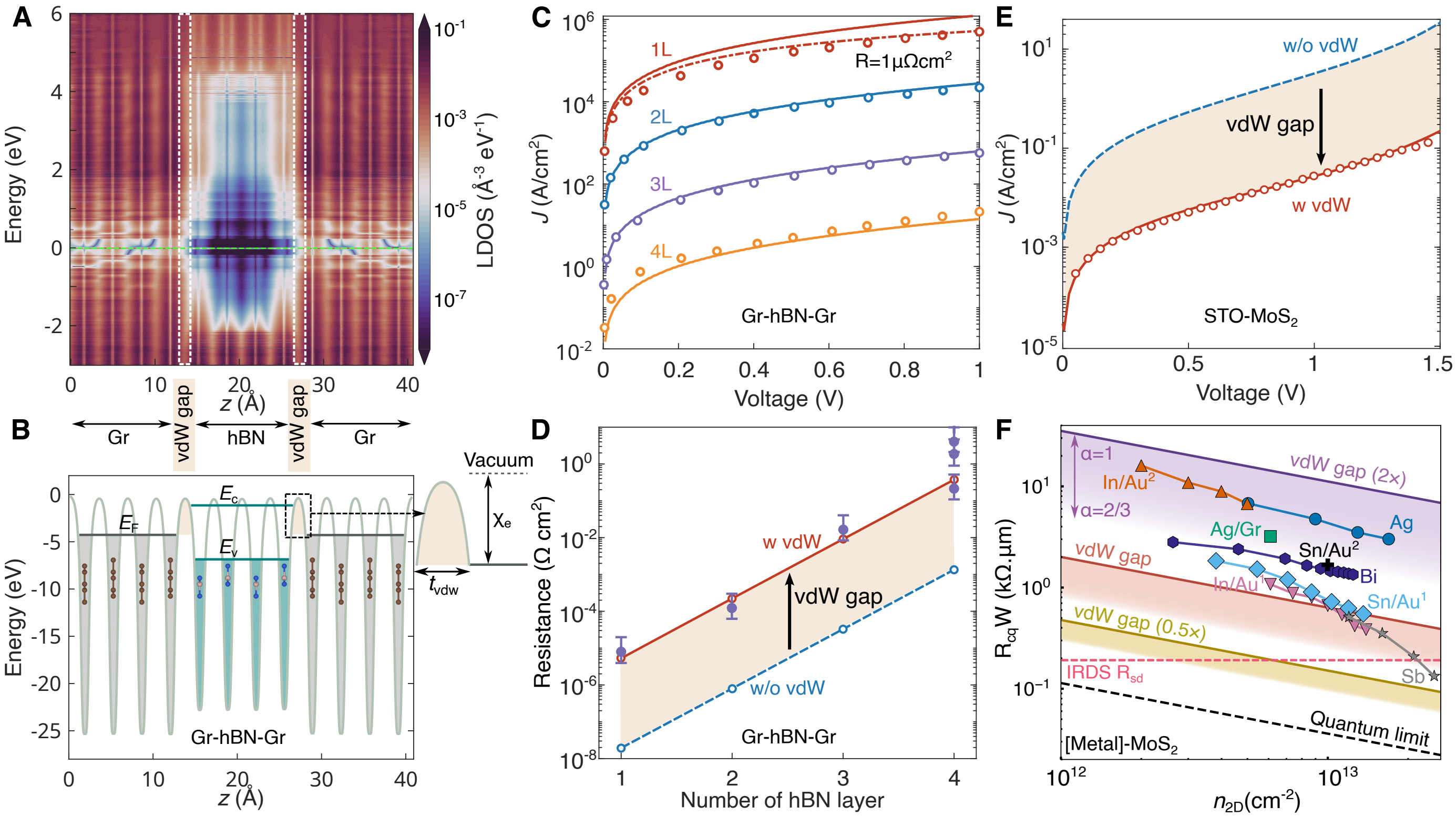

**Figure 4**: **Tunneling suppression caused by vdW gaps. (A)** Local density of states (LDOS) from ab initio NEGF simulations of a graphene–hBN–graphene tunnel junction under bias, showing strong decay within hBN and sharp drops at the graphene–hBN interfaces. **(B)** Planar-averaged electrostatic potential for the same structure, with vdW gap regions approaching the vacuum level. **(C)** Current density–voltage ($J$–$V$) characteristics for graphene–hBN–graphene tunneling junctions with one to four hBN layers. Solid lines denote simulations and symbols denote experimental data from Ref. (*17*). A small deviation for monolayer hBN at high bias is corrected by including a series resistance (dashed line). **(D)** Small-bias tunneling resistance as a function of hBN thickness. Including the vdW gaps (red) increases resistance by ∼260× compared with models that neglect the gap (blue), consistent with experiment (*17*). **(E)** Calculated gate leakage current for monolayer $MoS_2$ with an STO gate dielectric. Inclusion of a vdW gap of ∼1.4 Å suppresses tunneling by approximately an order of magnitude, in agreement with experiment (*18*). **(F)** Quantum contact resistivity of metal–$MoS_2$ vdW contacts as a function of vdW gap thickness (Section S6). Experimental data indicate vdW gaps of ∼ 1.2–2.7 Å (Ag (*20*), Ag/Gr (*21*), In/Au[1] and Sn/Au[1] (*22*), In/Au[2] (*23*), Sn/Au[2] (*24*), Sb (*25*), Bi (*26*)). The dashed line indicates the IRDS target for total source–drain contact resistance.

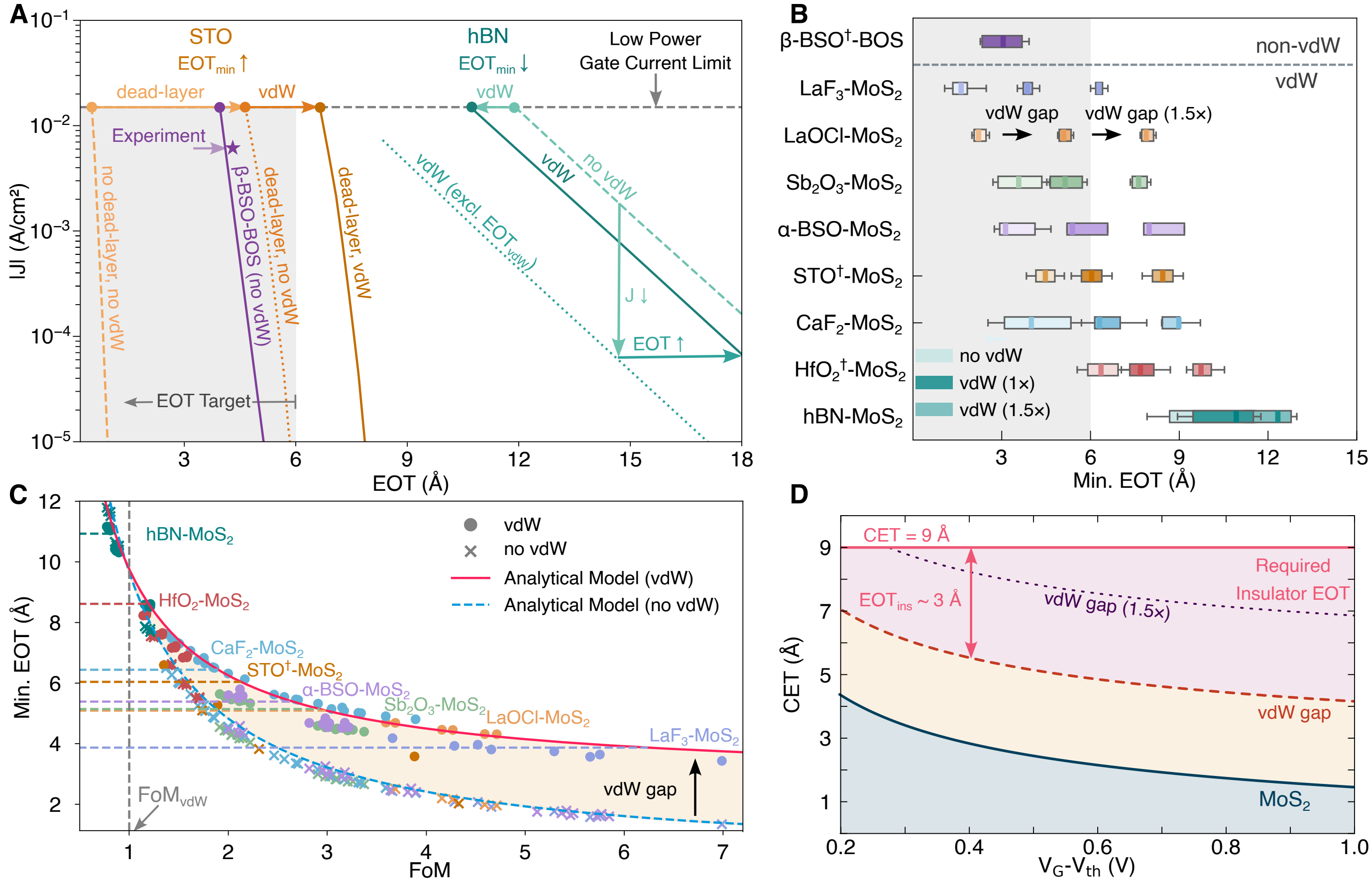

**Figure 5**: **Impact of vdW gaps on scaling limits.** **(A)** Calculated leakage current density versus EOT for hBN, STO, and $\beta$-BSO–BOS on $MoS_2$. For hBN, the vdW gap suppresses leakage and slightly reduces the minimum EOT, whereas for STO, interfacial dead layers and the vdW gap increase the minimum EOT to ~7 Å, placing it slightly above the IRDS scaling limit. In contrast, $\beta$-BSO–BOS forms a bonded interface without a vdW gap and maintains low leakage at small EOT. **(B)** Minimum achievable EOT for various insulators on $MoS_2$, comparing cases with no vdW gap, a DFT-predicted gap, and a more realistic enlarged gap (1.5×). Inclusion of the vdW gap increases the total EOT for most materials; only hBN shows a net benefit from leakage suppression. $\beta$-BSO–BOS achieves the target by avoiding the vdW-gap penalty. The total EOT must remain below 6 Å to meet IRDS CET targets. Insulators marked with a dagger (†) use a thickness-dependent permittivity model. **(C)** Minimum EOT versus insulator FoM. Curves show analytical predictions, and symbols denote numerical Tsu–Esaki results. For FoM $\lesssim$ 1 ($SiO_2$-like), the vdW gap improves scaling, whereas for larger FoM it limits EOT reduction. **(D)** $MoS_2$ CET versus $V_G - V_{th}$ for no vdW gap, a DFT-predicted gap, and an enlarged gap (1.5×). For the IRDS target CETtotal = 9 Å, CETMoS$_2$ ≈ 3 Å at 0.4 V leaves ~6 Å for the combined insulator and vdW-gap EOT.

# Acknowledgments

**Funding:** We acknowledge support by the Austrian Science Fund (FWF) and the European Research Council (ERC) under Grant agreement no. 101055379 (F2GO). The computational results presented have been achieved using the Vienna Scientific Cluster (VSC). We acknowledge fruitful discussions with Deji Akinwande, Mario Lanza, Aftab Nazir, Theresia Knobloch, and Dominic Waldhoer.

**Author contributions:** M.P. developed the methodology, performed all simulations, analyzed the data, interpreted the results, prepared the figures, and wrote the manuscript with feedback and suggestions from T.G. T.G. conceived the initial idea of the study, which was continuously refined in daily discussions with M.P. Both authors approved the final version of the manuscript.

**Competing interests:** There are no competing interests to declare.

**Data and materials availability:** All data and code needed to evaluate the conclusions in the paper are present in the paper and the Supplementary Materials. Additional datasets and analysis scripts are available at (*42*).

## Supplementary materials

Materials and Methods

Supplementary Text

Figs. S1 to S11

Tables S1 to S3

References *(43-62)*

# Supplementary Materials for

# Device-scaling constraints imposed by the van der Waals gap formed in two-dimensional materials

Mahdi Pourfath* and Tibor Grasser*

*Corresponding author. Email: pourfath@iue.tuwien.ac.at and grasser@tuwien.ac.at

**This PDF file includes:**

Materials and Methods

Supplementary Text

Figures S1 to S11

Tables S1 to S3

## Materials and Methods

**vdW gap estimation from atomic radii**:

As an independent and chemistry-agnostic estimate of vdW gap thicknesses, we evaluated gaps using tabulated atomic covalent and crystal van der Waals radii. For each element, a single-atom vdW contribution was defined as $\Delta_i = r_{\mathrm{vdW}} - r_{\mathrm{cov}}$, where covalent radii $r_{\mathrm{cov}}$ were taken from Ref. (*43*) and crystal vdW radii $r_{\mathrm{vdW}}$ from Ref. (*44*). Binary vdW gaps were then constructed as $\Delta_{ij} = \Delta_i + \Delta_j$ by enumerating unique elemental pairs. Statistical analysis over the resulting set yields a characteristic vdW gap scale that is independent of specific bonding chemistry. The underlying atomic radii data and derived distributions are summarized in Table S1 and visualized in Fig. S2. The resulting vdW-gap distribution is presented in Fig. 2D.

**DFT calculations**:

All first-principles calculations were performed using the Vienna *Ab initio* Simulation Package (VASP) (*45, 46*). Exchange–correlation effects were treated within the generalized gradient approximation (GGA) using the Perdew–Burke–Ernzerhof (PBE) functional, with van der Waals interactions accounted for via the DFT-D3 method of Grimme (*47*). A plane-wave energy cutoff of at least 580 eV was employed in all calculations.

For $MoS_2$–insulator stacks, heterostructures were constructed by applying biaxial strain below 1.5% to the insulating substrate to achieve in-plane commensurability with $MoS_2$, thereby minimizing strain-induced variations in the insulator band gap. In contrast, for Au–$MoS_2$ devices (Section S7), in-plane commensurability was ensured by applying a small biaxial strain ($< 1.5\%$) to Au while keeping $MoS_2$ unstrained to avoid band-gap modifications in the channel. For the STO–$MoS_2$ interface, bulk $SrTiO_3$ (001) slabs were cleaved with $TiO_2$ termination at the $MoS_2$ interface and SrO termination at the opposite surface; this asymmetric slab construction preserves stoichiometry. Both terminations are experimentally accessible; however, we focus on the $TiO_2|MoS_2$ interface owing to its experimental relevance and favorable band alignment for device applications.

All heterostructures were structurally relaxed along the direction perpendicular to the interface to determine the equilibrium interlayer spacing. Atomic positions were optimized until residual Hellmann–Feynman forces were below $1 \times 10^{-2}$ eV/Å, and electronic self-consistency was achieved using an energy convergence threshold of $10^{-8}$ eV. Brillouin-zone sampling during geometry relaxations employed Monkhorst–Pack **k**-point meshes with a density of approximately 0.04 $Å^{-1}$. For total energy and charge-density calculations, denser **k**-point meshes were used, with $\Gamma$-centered grids corresponding to a reciprocal-space spacing of approximately 0.03 $Å^{-1}$. A vacuum region of at least 30 Å was introduced along the out-of-plane ($z$) direction to eliminate spurious interactions between periodic images, which is particularly important for accurate evaluation of dielectric properties.

**Permittivity calculations**:

Permittivities were computed using both macroscopic and microscopic approaches. In-plane permittivities were obtained using macroscopic methods, as described in Section S1. For out-of-plane components, a microscopic analysis was employed to mitigate numerical artifacts associated with series-capacitance models and finite vacuum separations.

In these simulations, a small external electric field was applied perpendicular to the heterostructure (along the $z$ axis). The field-induced charge density was obtained by subtracting the zero-field electron density from the finite-field result. From this induced charge distribution, the local polarization was computed and used to construct an effective, position-dependent permittivity profile, $\kappa(z)$, which provides a compact representation of spatial variations in dielectric screening across the interface.

We emphasize that $\kappa(z)$ is an effective permittivity profile inferred from first-principles polarization under small applied fields, rather than a directly measurable macroscopic quantity. Device-level quantities, including capacitance, equivalent oxide thickness (EOT), and region-averaged permittivities, were obtained by spatially averaging the inverse permittivity profile, $1/\kappa(z)$, over the corresponding regions of the heterostructure.

The effective permittivity of the vdW gap, $\kappa_{\mathrm{vdW}}^{\mathrm{eff}}$, was extracted for each heterostructure by treating the spatially varying permittivity profile, $\kappa(z)$, as a series capacitance across the interfacial region. Specifically, the inverse permittivity, $1/\kappa(z)$, was spatially averaged over the vdW gap region to obtain an effective interfacial permittivity, as defined in Eq. (Eq. S16).

This procedure was applied consistently across a set of representative vdW heterostructures spanning different channel materials, gate dielectrics, and interface chemistries (Table Table S2). The characteristic vdW-gap permittivity, $\overline{\kappa}_{\mathrm{vdW}}$, reported in the main text was obtained by averaging the extracted $\kappa_{\mathrm{vdW}}^{\mathrm{eff}}$ values across this set, see Section S1.

**Layered-crystal permittivity model**:

To describe the thickness dependence of the out-of-plane permittivity in layered materials, we model multilayer crystals as series combinations of high-permittivity atomic layers and low-permittivity vdW gaps. In this framework, atomic layers are treated as dielectric slabs with relatively large permittivity, while vdW gaps between adjacent layers act as low-permittivity interfacial regions. The effective out-of-plane permittivity is determined by the series combination of these constituents and therefore depends on both the number of layers and the equilibrium interlayer spacing.

For an $N$-layer $MoS_2$ stack, the structure is represented as two surface half-layers and $(N - 1)$ internal layer–gap units arranged in series. This construction captures the reduction of permittivity in few-layer systems and its convergence toward the bulk limit as $N \to \infty$, in agreement with first-principles calculations and experimental observations.

Interfacial effects enter the model through the equilibrium vdW gap thickness, $t_{\mathrm{vdW}}$, which controls the relative contribution of low-permittivity regions to the overall series capacitance. Variations in $t_{\mathrm{vdW}}$, such as those induced by strain or pressure, modify the effective permittivity by changing the weight of the interlayer gaps. Analytical expressions and derivations of this model are provided in Section S2.

**vdW gap permittivity model**:

Equilibrium vdW gap thicknesses, $t_{\mathrm{vdW}}$, were obtained from density functional theory (DFT)–relaxed geometries of insulator–$MoS_2$ heterostructures. The vdW gap was defined as the separation between the terminal atomic planes of the insulator and the $MoS_2$ channel following structural relaxation. Across the set of representative systems considered in this work, the equilibrium gap thickness varied within a narrow range, typically within $\pm 0.2$ Å around the mean value.

For each heterostructure, the effective vdW-gap permittivity, $\kappa_{\mathrm{vdW}}^{\mathrm{eff}}$, was extracted from the spatially varying permittivity profile, $\kappa(z)$, by treating the vdW gap as a series capacitance. Specifically, the inverse permittivity, $1/\kappa(z)$, was spatially averaged over the vdW gap region to obtain an effective interfacial permittivity, as defined in Eq. (Eq. S6). The corresponding equivalent oxide thickness (EOT) contribution of the vdW gap was computed using the standard definition $\mathrm{EOT}_{\mathrm{vdW}} = (\kappa_{\mathrm{SiO_2}}/\kappa_{\mathrm{vdW}}^{\mathrm{eff}})t_{\mathrm{vdW}}$, where $\kappa_{\mathrm{SiO_2}} = 3.9$.

The dependence of the effective vdW-gap permittivity on the equilibrium gap thickness was captured using a simple analytical model,

$$\kappa_{\mathrm{vdW}}^{-1} = 1 - \frac{c}{t_{\mathrm{vdW}}}, \tag{S1}$$

where $c$ is a fitting parameter that accounts for interfacial polarization associated with evanescent electronic tails at the interface. The parameter $c$ was extracted by fitting the model to the first-principles data across the set of representative heterostructures shown in Fig. 3D,E and summarized in Table Table S2. The average fitted value was $\overline{c} = 0.72$ Å.

Using this model and the extracted gap thicknesses, the average effective vdW-gap permittivity across the considered systems was $\overline{\kappa}_{\mathrm{vdW}} \approx 2$, corresponding to an average vdW-gap contribution to the equivalent oxide thickness of $\overline{\mathrm{EOT}}_{\mathrm{vdW}} \approx 2.7$ Å. These values are reported in Table Table S2.

**$MoS_2$ Capacitance-equivalent thickness calculations**:

For the extraction of the intrinsic $MoS_2$ channel contribution to CET, hybrid density-functional calculations were performed for slab geometries using the HSE06 screened-exchange functional. A plane-wave cutoff of 520 eV was employed with PAW potentials, non-spherical gradient corrections, and augmented FFT grids to ensure accurate electrostatics and charge-density profiles. A Γ-centered $26 \times 26 \times 1$ **k**-point mesh was used to converge the quantum capacitance and dielectric response of monolayer $MoS_2$. Dipole corrections were applied along the out-of-plane direction to eliminate spurious periodic interactions. Spatially resolved charge and local density-of-states (LDOS) distributions were obtained from partial charge-density projections and post-processed using the DensityTool package (*48*). These LDOS-derived charge distributions were used to extract bias-dependent charge centroids and intrinsic channel capacitances, which serve as the baseline for the CET scaling analysis. Additional computational details are provided in Section S8.

**Tunneling models and simulations**:

Tunneling across vdW gaps was analyzed using a combination of compact analytical models and first-principles quantum transport simulations. Analytical screening of tunneling currents was performed using a semiclassical WKB formulation within the Tsu–Esaki framework, employing literature-based material parameters and experimentally anchored band alignments. This approach provides physical transparency and enables systematic evaluation of vdW-gap effects across a wide range of materials and device geometries.

Within the WKB formulation, tunneling through the vdW gap was treated as transmission through a vacuum-like barrier. Because the gap lacks crystal periodicity, electrons traversing the vdW region were modeled using the free-electron mass $m_0$. Deviations of the interfacial potential from an ideal rectangular profile were captured using a dimensionless barrier-shape factor $\alpha$, which effectively accounts for the spatial variation of the barrier under realistic electrostatics. In the rectangular-barrier limit, $\alpha = 1$, whereas smoother, trapezoidal barriers can reduce $\alpha$ toward $2/3$. In practice, $\alpha$ increases toward unity for larger vdW gaps (more rectangular profiles) and decreases for narrower gaps, where neighboring atomic potentials soften the barrier and lower the effective action. A representative range of $\alpha$ values was therefore considered, with $\alpha \approx 2/3$ used to describe softer, trapezoidal barriers at smaller or perturbed gaps and $\alpha \to 1$ corresponding to the rectangular-barrier limit expected for larger vdW gaps. The sensitivity of the tunneling and contact resistance to both $\alpha$ and $t_{\mathrm{vdW}}$ was explicitly evaluated in the parameter sweeps shown in Fig. 4F and described in Section S6.

To validate the analytical description and to capture atomic-scale electrostatics at vdW interfaces, representative junctions were simulated using density-functional theory coupled with the nonequilibrium Green's function (DFT–NEGF) formalism. These simulations explicitly resolve the electrostatic potential profile, local density of states, and transmission across vdW heterostructures. Detailed descriptions of the NEGF methodology, benchmarking against analytical models, and convergence tests are provided in Section S7. The extension of this tunneling framework to metal–2D semiconductor contacts and quantum contact resistance is described in Section S6.

For the graphene–hBN–graphene junction modeling, we adopted literature-based parameters that reproduce experimentally observed characteristics of similar vertical heterostructures. The effective electron mass in hBN was taken as $m^* = 0.5\,m_0$ (*17*), and the work function of graphene was set to $\Phi_{\mathrm{G}} = 4.5$ eV for undoped graphene (*16*). To account for unintentional environmental and fabrication-induced doping, a small Fermi-level shift of approximately $\Delta E_{\mathrm{F}} \approx 0.1$ eV was included, corresponding to carrier densities of $n \sim 10^{12}$–$10^{13}$ cm$^{-2}$. Parameter sweeps confirmed that varying the residual doping within the range $\Delta E_{\mathrm{F}} \approx 0.05$–$0.15$ eV does not greatly alter the calculated current–voltage characteristics or the qualitative tunneling trends.

The electron affinity of hBN, $\chi_{\mathrm{hBN}}$, is known to depend sensitively on surface termination, thickness, and preparation conditions. Ultraviolet photoelectron spectroscopy measurements on APCVD-grown and commercial multilayer hBN report values of $\chi_{\mathrm{hBN}}$ between approximately 1.6 and 2.4 eV (*49*). We therefore adopted an effective value of $\chi_{\mathrm{hBN}} = 2.0$ eV for the simulations, which lies well within the experimentally observed range.

Device-level quantum transport calculations were performed using density-functional theory combined with the nonequilibrium Green's function (DFT–NEGF) formalism (*50*), as implemented in QuantumATK (*51*). The calculations employed the Perdew–Burke–Ernzerhof (PBE) exchange–correlation functional and norm-conserving SG15 Medium LCAO basis sets (*52*). A real-space density mesh cutoff of 150 Ry was used. $k$-point sampling for the self-consistent device calculations employed a density of 4.0 $\text{Å}^{-1}$ along each transverse direction and 150.0 $\text{Å}^{-1}$ along the transport direction. Transverse Brillouin-zone integrations for the transmission and the local density of states employed a $k$-point density of 10.0 $\text{Å}^{-1}$ in each in-plane direction. The total-energy convergence threshold was better than $10^{-6}$ eV.

Regarding the accuracy and computational cost of NEGF–DFT, we note that standard GGA functionals such as PBE underestimate bulk band gaps, which can greatly affect tunneling probabilities. Although hybrid functionals (e.g., HSE06) correct bulk gaps, they are prohibitively expensive for realistic vdW heterostructures in NEGF, where large supercells are required to minimize interfacial strain, and they often require tuning of the exact-exchange fraction to match experimental gaps. While such tuning may be acceptable for a single bulk insulator, transferring the same parameter to a heterostructure can overcorrect the partner material and yield erroneous band offsets. Accordingly, DFT–NEGF was used primarily for representative validation cases and to obtain relaxed geometries and qualitative trends, whereas device-level screening relied on a compact Tsu–Esaki formulation employing experimentally anchored material parameters. This combined approach has been benchmarked against both ab initio NEGF and multiple experimental datasets (see Fig. 4 and Section S7).

The tunneling calculations employed three primary material parameters for each insulating layer: the relative permittivity $\kappa$, the electron affinity $\chi$, and the tunneling effective mass $m_{\mathrm{t}}^{*}/m_0$, all taken from the literature. Thickness-dependent permittivity effects were included only where experimentally justified; all other materials were treated using thickness-independent bulk permittivities. Band offsets in insulator–$MoS_2$ stacks were obtained using $\chi \approx 4.3$ eV and $E_{\mathrm{g}} \approx 2.0$ eV for $MoS_2$, and $\chi \approx 4.0$ eV with $E_{\mathrm{g}} = 1.0$ eV for BOS. The complete set of material parameters used in the tunneling calculations is summarized in Table S3.

## Supplementary Text

### S1 vdW gap permittivity

**Note:** Throughout this section, all permittivities refer to the *electronic* contribution, $\kappa^{\infty}$, which describes the system's response to an external electric field via electronic charge redistribution, as computed by linear-response DFT or field-induced polarization. The superscript $\infty$ is omitted for simplicity. The dielectric response of the vdW gap reflects electronic polarization, as the region contains no atoms and no lattice contribution. We use $\kappa(z)$ to denote an effective permittivity profile constructed from the induced polarization; this is a modeling construct to visualize screening, and macroscopic quantitity are obtained by region averaging as specified below.

**Macroscopic approach**: Using DFT calculations combined with classical electrostatics, the out-of-plane ($\kappa^{\perp}_{\mathrm{vdW}}$) and in-plane ($\kappa^{\parallel}_{\mathrm{vdW}}$) permittivities of the vdW gap are derived. The effective dielectric response of composite stacks under a small homogeneous electric field is computed using the macroscopic dielectric framework of Ref. (*53*), based on perturbative linear-response DFT.

As illustrated in Fig. S7, a simulation cell comprising an insulator, a vdW gap, and a monolayer of $MoS_2$ is constructed to calculate the effective dielectric response of the entire supercell. The system is embedded in at least 20 Å of vacuum to avoid spurious interactions with periodic images imposed by boundary conditions. The decomposition of the total effective dielectric response into contributions from the insulator, $MoS_2$, and the vdW gap is not directly accessible from DFT. Instead, it is inferred using classical electrostatics by modeling the system as a combination of capacitors in series or parallel (*54*), depending on the direction of the applied electric field. This approximation is justified by the spatial separation between components and the non-overlapping nature of their wavefunctions in typical slab geometries. Intrinsic dielectric properties of each region are determined from three separate DFT calculations within a common supercell of length $L_{\mathrm{cell}}$, all incorporating a vacuum region: (i) the combined system comprising the insulator, vdW gap, and monolayer $MoS_2$ (yielding $\kappa_{\mathrm{c,Ins\text{-}MoS_2}}$); (ii) the isolated insulator slab ($\kappa_{c,\mathrm{ins}}$); and (iii) the isolated $MoS_2$ layer ($\kappa_{\mathrm{c,MoS_2}}$). In each case, $\kappa_{\mathrm{c}}$ represents the effective permittivity of the entire cell, including vacuum, whereas $\kappa_{\mathrm{ins}}$, $\kappa_{\mathrm{MoS_2}}$, and $\kappa_{\mathrm{vdW}}$ denote the intrinsic permittivities of the respective material regions. The intrinsic value $\kappa_{\mathrm{vdW}}$ is then extracted by combining the computed effective quantitities through classical electrostatics.

**Out-of-plane direction**: The permittivity of each configuration is derived from the inverse-capacitance (series capacitor) model. For the *combined system* one can write:

$$\frac{L_{\text{cell}}}{\kappa^{\perp}_{\text{c,Ins–MoS}_2}} = \frac{L_{\text{ins}}}{\kappa^{\perp}_{\text{ins}}} + \frac{L_{\text{MoS}_2}}{\kappa^{\perp}_{\text{MoS}_2}} + \frac{L_{\text{vdW}}}{\kappa^{\perp}_{\text{vdW}}} + \frac{L_{\text{vac}}}{1} \tag{S2}$$

Using $L_{\text{vac}} = L_{\text{cell}} - L_{\text{ins}} - L_{\text{MoS}_2} - L_{\text{vdW}}$, one gets:

$$L_{\text{cell}}\left(\frac{1}{\kappa^{\perp}_{\text{c,Ins–MoS}_2}} - 1\right) = L_{\text{ins}}\left(\frac{1}{\kappa^{\perp}_{\text{ins}}} - 1\right) + L_{\text{MoS}_2}\left(\frac{1}{\kappa^{\perp}_{\text{MoS}_2}} - 1\right) + L_{\text{vdW}}\left(\frac{1}{\kappa^{\perp}_{\text{vdW}}} - 1\right) \tag{S3}$$

For the isolated insulator and isolated $MoS_2$, respectively, one obtains:

$$L_{\text{cell}}\left(\frac{1}{\kappa^{\perp}_{\text{c,ins}}} - 1\right) = L_{\text{ins}}\left(\frac{1}{\kappa^{\perp}_{\text{ins}}} - 1\right), \qquad L_{\text{cell}}\left(\frac{1}{\kappa^{\perp}_{\text{c,MoS}_2}} - 1\right) = L_{\text{MoS}_2}\left(\frac{1}{\kappa^{\perp}_{\text{MoS}_2}} - 1\right) \tag{S4}$$

Substituting Eq. S3–Eq. S4 into Eq. S2:

$$L_{\text{cell}}\left(\frac{1}{\kappa^{\perp}_{\text{c,Ins–MoS}_2}} - \frac{1}{\kappa^{\perp}_{\text{c,ins}}} - \frac{1}{\kappa^{\perp}_{\text{c,MoS}_2}} + 1\right) = L_{\text{vdW}}\left(\frac{1}{\kappa^{\perp}_{\text{vdW}}} - 1\right) \tag{S5}$$

Solving for $\kappa^{\perp}_{\text{vdW}}$:

$$\boxed{\kappa^{\perp}_{\text{vdW}} = \frac{1}{1 + \frac{L_{\text{cell}}}{L_{\text{vdW}}}\left(\frac{1}{\kappa^{\perp}_{\text{c,Ins–MoS}_2}} - \frac{1}{\kappa^{\perp}_{\text{c,ins}}} - \frac{1}{\kappa^{\perp}_{\text{c,MoS}_2}} + 1\right)}} \tag{S6}$$

**In-plane direction**: The in-plane response follows a volume-averaged (parallel capacitor) model. For the combined system, one obtains:

$$L_{\text{cell}}\kappa^{\parallel}_{\text{c,Ins–MoS}_2} = L_{\text{ins}}\kappa^{\parallel}_{\text{ins}} + L_{\text{MoS}_2}\kappa^{\parallel}_{\text{MoS}_2} + L_{\text{vdW}}\kappa^{\parallel}_{\text{vdW}} + L_{\text{vac}} \tag{S7}$$

Using $L_{\text{vac}} = L_{\text{cell}} - L_{\text{ins}} - L_{\text{MoS}_2} - L_{\text{vdW}}$:

$$L_{\text{cell}}\left(\kappa^{\parallel}_{\text{c,Ins–MoS}_2} - 1\right) = L_{\text{ins}}\left(\kappa^{\parallel}_{\text{ins}} - 1\right) + L_{\text{MoS}_2}\left(\kappa^{\parallel}_{\text{MoS}_2} - 1\right) + L_{\text{vdW}}\left(\kappa^{\parallel}_{\text{vdW}} - 1\right) \tag{S8}$$

Similarly, for the isolated cases:

$$L_{\text{cell}}\left(\kappa^{\parallel}_{\text{c,ins}} - 1\right) = L_{\text{ins}}\left(\kappa^{\parallel}_{\text{ins}} - 1\right), \quad L_{\text{cell}}\left(\kappa^{\parallel}_{\text{c,MoS}_2} - 1\right) = L_{\text{MoS}_2}\left(\kappa^{\parallel}_{\text{MoS}_2} - 1\right) \tag{S9}$$

Substituting into the expression yields:

$$L_{\text{cell}}\left(\kappa^{\parallel}_{\text{c,Ins–MoS}_2} - \kappa^{\parallel}_{\text{c,ins}} - \kappa^{\parallel}_{\text{c,MoS}_2} + 1\right) = L_{\text{vdW}}\left(\kappa^{\parallel}_{\text{vdW}} - 1\right) \tag{S10}$$

Solving for $\kappa^{\|}_{\mathrm{vdW}}$:

$$\boxed{\kappa^{\|}_{\mathrm{vdW}} = 1 + \frac{L_{\mathrm{cell}}}{L_{\mathrm{vdW}}}\left(\kappa^{\|}_{\mathrm{c,Ins\text{-}MoS_2}} - \kappa^{\|}_{\mathrm{c,ins}} - \kappa^{\|}_{\mathrm{c,MoS_2}} + 1\right)} \tag{S11}$$

**Microscopic approach**: This section outlines the microscopic procedure used to extract the permittivity of a vdW gap from first-principles calculations. It complements the macroscopic perspective presented next by analyzing the spatial distribution of the dielectric response. The method involves evaluating the charge redistribution induced by a small external electric field applied along the out-of-plane ($z$) direction, as well as computing the resulting polarization profiles and effective electric fields. This approach provides local insights into the dielectric behavior across the slab (*55, 56*).

A homogeneous electric field $E_{\mathrm{ext}}$ is applied along the $z$-axis (Fig. S7), inducing a redistribution of electronic charge. The corresponding difference in charge density is obtained as:

$$\rho_{\mathrm{ind}}(z) = \rho^{E}(z) - \rho^{0}(z), \tag{S12}$$

where $\rho^{E}(z)$ and $\rho^{0}(z)$ are the planar-averaged charge densities with and without the external electric field, respectively. The spatially varying polarization $P(z)$ is then computed by integrating the induced charge density:

$$P(z) = -\int_{z_0}^{z} \rho_{\mathrm{ind}}(z')\,\mathrm{d}z'. \tag{S13}$$

The corresponding induced screening field $E_{\mathrm{ind}}(z)$ and the total effective electric field $E_{\mathrm{eff}}(z)$ are:

$$E_{\mathrm{ind}}(z) = -\frac{1}{\varepsilon_0}P(z), \qquad E_{\mathrm{eff}}(z) = E_{\mathrm{ext}} + E_{\mathrm{ind}}(z). \tag{S14}$$

The effective permittivity profile at each position $z$ is then given by:

$$\kappa(z) = \frac{E_{\mathrm{ext}}}{E_{\mathrm{eff}}(z)} = 1 + \frac{P(z)}{\varepsilon_0 E_{\mathrm{eff}}(z)}. \tag{S15}$$

To mitigate atomic-scale oscillations in the permittivity profile, a Gaussian filter can be applied to $\rho_{\mathrm{ind}}(z)$ and the derived quantities. The effective out-of-plane permittivity of the vdW gap is then computed as:

$$\boxed{\frac{1}{\kappa^{\perp}_{\mathrm{vdW}}} = \frac{1}{L_{\mathrm{vdW}}}\int_{z_1}^{z_2} \frac{1}{\kappa(z)}\,\mathrm{d}z.} \tag{S16}$$

Only the region-averaged value in Eq. S16 is used to map this profile to series capacitance; the pointwise $\kappa(z)$ is not interpreted as a layer-resolved material constant.

## S2 Permittivity thickness dependence and interfacial effects

Dielectric materials at the nanoscale often deviate from bulk behavior due to finite-size effects, interfacial phenomena, and structural inhomogeneities. This section focuses on two key cases and the models describing them: (i) the layer-dependent permittivity observed in vdW materials, and (ii) dead layer effects typically associated with interfaces in high-$\kappa$ dielectrics but also arising from grain boundaries, lattice defects, or other microstructural features. Both phenomena highlight the critical role of structural and interfacial effects in shaping the dielectric properties of nanoscale devices.

**Thickness-Dependent Permittivity in vdW Materials:** vdW materials display distinctive dielectric behavior due to their layered structure and low-permittivity gaps between atomic planes. The effective permittivity often varies with the number of layers, especially in thin films, reflecting differences in polarization between interfacial and inner layers. This variation can be modeled as a series of dielectric regions: the atomic layers with similar permittivity, and the vdW gaps and interfacial layers, which differ from bulk properties. In ultrathin vdW structures, surface and interface contributions play a dominant role, lowering the effective permittivity relative to bulk values. As the number of layers increases, this influence diminishes, and the permittivity approaches its bulk limit. This behavior is conceptually similar to the dead layer effect in thin films, where interfacial regions with reduced permittivity decrease the total capacitance.

The use of a series capacitance model to describe the out-of-plane permittivity in a layered vdW stack is illustrated in Fig. S5A, where the total thickness of an $N$-layer stack is given by: $t_{\text{total}} = t_{\text{SL}} + (N-1)\,(t_{\text{slab}} + t_{\text{vdW}})$, where $t_{\text{SL}}$ is the effective thickness of a single layer, $t_{\text{slab}}$ the dielectric thickness near atomic planes, and $t_{\text{vdW}}$ the vdW gap thickness. A common approximation is: $t_{\text{SL}} = t_{\text{slab}} + t_{\text{vdW}}$. The effective permittivity $\kappa_{\text{eff}}$ follows:

$$\frac{t_{\text{total}}}{\kappa_{\text{eff}}} = \frac{t_{\text{SL}}}{\kappa_{\text{SL}}} + (N-1)\left(\frac{t_{\text{slab}}}{\kappa_{\text{slab}}} + \frac{t_{\text{vdW}}}{\kappa_{\text{vdW}}}\right), \tag{S17}$$

where $\kappa_{\text{SL}}$, $\kappa_{\text{slab}}$, and $\kappa_{\text{vdW}}$ are the permittivities of the surface-influenced layer, the atomic-layer region, and the vdW gap, respectively. As $N$ increases, interfacial effects become negligible, and $\kappa_{\text{eff}}$ approaches the series combination of $\kappa_{\text{slab}}$ and $\kappa_{\text{vdW}}$. Moreover, external factors such as strain can reduce the vdW gap thickness, enhancing polarization and increasing the permittivity, as shown in Fig. 3F of the manuscript.

**Dead layer effects in thin-film dielectrics:** Interfacial dead layers commonly arise in high-$\kappa$ dielectrics, particularly when integrated into ultrathin gate stacks. These regions near interfaces exhibit suppressed polarization and lower permittivity compared to the bulk (*6, 57*), as illustrated in Fig. S5B, leading to a degradation in effective dielectric response and reduced capacitance. Dead layers can also result from microstructural inhomogeneities such as grain boundaries, dislocations, or defects, which locally suppress dielectric screening. Although the focus here is on interfacial dead layers, other types can be similarly treated using a series capacitance model.

The overall stack is modeled as two low-permittivity interfacial dead layers of thickness $t_{\mathrm{DL}}$ (one on each side), sandwiching a central bulk region of thickness $t_{\mathrm{bulk}}$ and permittivity $\kappa_{\mathrm{bulk}}$. The effective dielectric response of the entire stack is determined by the series combination of the bulk and dead layer contributions. The effective permittivity of the dead layer is obtained from a graded inverse-permittivity profile. As demonstrated in *ab-initio* calculations (*6*), the inverse permittivity $\kappa^{-1}(z)$ within each dead layer can be assumed to vary linearly from vacuum ($\kappa = 1$ at $z = 0$) to the bulk value ($\kappa = \kappa_{\mathrm{bulk}}$ at $z = t_{\mathrm{DL}}$). This assumption yields:

$$\kappa_{\mathrm{DL}}^{-1}(z) = \left(1 - \frac{z}{t_{\mathrm{DL}}}\right) + \frac{z}{t_{\mathrm{DL}}} \cdot \frac{1}{\kappa_{\mathrm{bulk}}}. \tag{S18}$$

The effective permittivity of the dead layer, $\kappa_{\mathrm{DL}}^{\mathrm{eff}}$, is then calculated by inverting the spatial average of $\kappa_{\mathrm{DL}}^{-1}(z)$:

$$\kappa_{\mathrm{DL}}^{\mathrm{eff}} = \left[\frac{1}{t_{\mathrm{DL}}} \int_0^{t_{\mathrm{DL}}} \kappa_{\mathrm{DL}}^{-1}(z)\, \mathrm{d}z\right]^{-1}. \tag{S19}$$

Evaluating the integral leads to:

$$\kappa_{\mathrm{DL}}^{\mathrm{eff}} = \left[\frac{1}{2}\left(1 + \frac{1}{\kappa_{\mathrm{bulk}}}\right)\right]^{-1} = \frac{2\kappa_{\mathrm{bulk}}}{1 + \kappa_{\mathrm{bulk}}}. \tag{S20}$$

This effect is modeled by treating the dielectric stack as a series combination of the bulk and interfacial regions. The total effective permittivity $\kappa_{\mathrm{tot}}^{\mathrm{eff}}$ is expressed as:

$$\frac{t}{\kappa_{\mathrm{tot}}^{\mathrm{eff}}} = \frac{t_{\mathrm{bulk}}}{\kappa_{\mathrm{bulk}}} + \frac{2t_{\mathrm{DL}}}{\kappa_{\mathrm{DL}}^{\mathrm{eff}}}, \tag{S21}$$

where $t = t_{\mathrm{bulk}} + 2t_{\mathrm{DL}}$ denotes the total thickness. This expression can be rearranged to isolate the dead layer contribution (*18*):

$$\boxed{\frac{t}{\kappa_{\mathrm{tot}}^{\mathrm{eff}}} = \frac{t}{\kappa_{\mathrm{bulk}}} + D.} \tag{S22}$$

The parameter $D$ is introduced to characterize the contribution of interfacial dead layers to the inverse capacitance per unit area in thin-film dielectrics. This formulation captures how regions with reduced permittivity near interfaces act in series with the bulk dielectric, thereby lowering the overall capacitance. As the total thickness $t$ is scaled down, the relative impact of the dead layer increases. Because the field-dependent reduction of permittivity is localized to the interfacial high-field region (*10, 28*), one can account for it by a slight, bias-dependent increase of the interfacial term, $D = D_0 + \Delta D(E)$, where $\Delta D(E)$ captures the field-induced reduction in permittivity.

This model has been successfully applied to extract dead layer characteristics from experimental data. For instance, in the case of $SrTiO_3$ (STO), a bulk permittivity of 270 and a dead layer thickness of 1.61 Å were reported (*18*). These values were extracted from metal–insulator–metal (MIM) capacitors, where the oxide is conformally deposited between two metal electrodes, minimizing the likelihood of interfacial vacuum gaps. Moreover, ab-initio simulations show that STO exhibits intrinsic interfacial suppression of permittivity due to soft-phonon and structural effects (*6*), particularly near metal interfaces. Therefore, additional corrections for vdW gaps are generally not required in MIM-based extractions of dielectric properties.

In contrast, the $HfO_2$ dataset analyzed here was obtained from transistor gate stacks, where the oxide interfaces a semiconductor channel. In such metal–insulator–semiconductor (MIS) structures (*32*), an interfacial vdW gap appears. This gap acts as a low-permittivity layer in series with the oxide and contributes to the total inverse areal capacitance. While its exact presence or thickness may vary across systems, accounting for such a contribution can improve the physical consistency of extracted parameters, particularly when modeling intrinsic dielectric behavior in ultrathin oxides.

In the experimental $HfO_2$ dataset used in Fig. S5C, the reported effective permittivity values were extracted based on electrical thickness measurements. Since the data were obtained from transistor gate stacks, we interpret the measured thickness as including both the physical oxide thickness $t$ and a finite interfacial vdW gap $t_{\mathrm{vdW}}$. This interpretation yields the relation

$$\frac{t + t_{\mathrm{vdW}}}{\kappa_{\mathrm{meas}}(t)} = \frac{t}{\kappa_{\mathrm{HfO_2,bulk}}} + D + \frac{t_{\mathrm{vdW}}}{\kappa_{\mathrm{vdW}}}, \tag{S23}$$

where $\kappa_{\mathrm{HfO_2,bulk}}$ is the bulk dielectric permittivity of $HfO_2$, $D$ is the dead layer degradation parameter introduced in Eq. S22, and $\kappa_{\mathrm{vdW}}$ is the effective permittivity of the vdW gap. Throughout this work we use $t_{\mathrm{vdW}} = 1.48$ Å and $\kappa_{\mathrm{vdW}} = 2.1$ (Table S2), corresponding to a capacitance-equivalent thickness (CET) interfacial contribution of $t_{\mathrm{vdW}}/\kappa_{\mathrm{vdW}} = 0.71$ Å.

To isolate the intrinsic dielectric response of $HfO_2$, the vdW contribution is removed in CET space. We therefore define an $HfO_2$-only effective permittivity by

$$\frac{t}{\kappa_{\mathrm{HfO_2,eff}}(t)} = \frac{t + t_{\mathrm{vdW}}}{\kappa_{\mathrm{meas}}(t)} - \frac{t_{\mathrm{vdW}}}{\kappa_{\mathrm{vdW}}}, \tag{S24}$$

such that the corrected data satisfy the standard dead layer relation

$$\frac{t}{\kappa_{\mathrm{HfO_2,eff}}(t)} = \frac{t}{\kappa_{\mathrm{HfO_2,bulk}}} + D. \tag{S25}$$

The corrected permittivity data $\kappa_{\mathrm{HfO_2,eff}}(t)$ were refit using Eq. S25, yielding a bulk permittivity of $\kappa_{\mathrm{HfO_2,bulk}} = 16.7$ and a dead layer degradation parameter $D = 0.69$ Å. By contrast, neglecting the vdW gap correction to the HfO2 permittivity values yielded $\kappa_{\mathrm{HfO_2,bulk}} = 16.2$ and a larger dead layer $D = 1.16$ Å.

This analysis highlights that vdW corrections are especially relevant for MIS structures where non-bonded interfaces can occur, and where interfacial effects may dominate in the sub-nanometer regime. By contrast, such corrections are generally less critical in MIM configurations, where bonding is more uniform.

## S3 Simplified Tsu–Esaki model for tunneling current

The Tsu–Esaki formalism provides the foundation for rigorously modeling electron tunneling through a dielectric stack under elastic transport conditions. This model evaluates the total current as an energy-integrated product of the quantum mechanical transmission probability and the thermally broadened supply function. The general expression for the current density reads:

$$J = \frac{q}{2\pi^2\hbar} \int \mathrm{d}E_z\, T(E_z) \left[ \frac{m^*}{\hbar^2} \int_0^\infty \left( f_\mathrm{M}(E_z + E_\perp) - f_\mathrm{ch}(E_z + E_\perp) \right) \mathrm{d}E_\perp \right], \tag{S26}$$

where $E = E_z + E_\perp$ is the total electron energy split into longitudinal and transverse components, and $f_\mathrm{M}$, $f_\mathrm{ch}$ are the Fermi–Dirac distribution functions in the metallic gate and the channel, respectively. $T(E_z)$ is the transmission probability at energy $E_z$.

In the following, we derive a simplified, closed-form expression to study the impact of key parameters. This analytical model is intended solely for physical insight and trend analysis and is not used to generate any of the numerical results shown in the figures. Assuming parabolic, isotropic bands, the transverse momentum integration can be carried out analytically, yielding the supply function:

$$N(E_z) = k_\mathrm{B} T \ln \left( \frac{1 + \exp\left[ (\phi_\mathrm{M} - E_z)/(k_\mathrm{B}T) \right]}{1 + \exp\left[ (E_\mathrm{F,ch} - E_z)/(k_\mathrm{B}T) \right]} \right)$$

where all energies are referenced to the vacuum level, $\phi_\mathrm{M}$ is the (negative) metal work function, and $E_\mathrm{F,ch}$ is the channel Fermi level. The total current density can then be written as

$$J = \frac{qm^*}{2\pi^2\hbar^3} \int T(E) N(E)\, \mathrm{d}E, \tag{S27}$$

where the subscript $z$ is omitted and $E$ denotes the longitudinal energy for simplicity.

In tunneling-dominated regimes, the supply function decays exponentially with increasing energy, so the dominant contribution to the current arises from states near the channel conduction-band edge,

$$E = E_\mathrm{c,ch} = \chi_\mathrm{ch} - qV_\mathrm{G}, \tag{S28}$$

where $\chi_\mathrm{ch}$ is the channel electron affinity and $V_\mathrm{G}$ is the applied gate voltage. Taking the gate as the electrostatic reference, the channel band edge shifts with $V_\mathrm{G}$.

During operation, the channel can become degenerate (e.g., $E_\mathrm{F,ch} - E_\mathrm{c,ch}$ can reach hundreds of meV), in which case a full Fermi–Dirac description is required. Nevertheless, to obtain a closed-form first-order approximation and expose the dependence on key parameters, we adopt the Maxwell–Boltzmann (MB) limit as an analytical ansatz:

$$N(E) \approx k_B T \exp\left(\frac{\phi_M - E}{k_B T}\right) = k_B T \exp\left(\frac{\phi_M - \chi_{ch} + qV_g}{k_B T}\right) \approx k_B T \exp\left(-\frac{\phi_B}{k_B T}\right) \exp\left(\frac{qV_g}{k_B T}\right). \tag{S29}$$

where the effective barrier height is given by

$$\phi_B = \chi_M - \phi_{ch}. \tag{S30}$$

This approximation is used only for trend analysis and is not used to generate any of the numerical results shown in the figures, which employ the full Fermi–Dirac supply function.

Assuming $T(E)$ is approximately constant over a thermal window $\Delta E \sim k_B T$, the current becomes: $J \approx \left(\frac{qm^*}{2\pi^2\hbar^3} k_B T\right) T(E) N(E)$. Thus,

$$\boxed{J \approx \underbrace{\frac{qm^*(k_B T)^2}{2\pi^2\hbar^3} \exp\left(-\frac{\phi_B}{k_B T}\right) \exp\left(\frac{qV_g}{k_B T}\right)}_{J_0} T(E_{c,ch})}, \tag{S31}$$

where the transmission probability is evaluated at the energy corresponding to the conduction band edge of the channel, as defined in Eq. S28. Using the trapezoidal barrier approximation, the transmission at this energy can be estimated as

$$T(E_{c,ch}) \approx \exp\left[-\frac{4\sqrt{2m^*}}{3\hbar q E_{ins}} \left(\left(\Delta E + qV_g\right)^{3/2} - \Delta E^{3/2}\right)\right], \quad \Delta E = \chi_{ch} - \chi_{ins}, \quad E_{ins} = \frac{V_g}{t_{ins}}, \tag{S32}$$

with $\Delta E$ representing the band offset between the conduction bands of the channel and the insulator, and $E_{ins} = V_g/t_{ins}$ denoting the electric field across the insulator of thickness $t_{ins}$. The prefactor in Eq. S31 resemble those of the classical Richardson equation for thermionic emission:

$$J = A^* T^2 \exp\left(-\frac{q\phi_B}{k_B T}\right), \quad \text{with} \quad A^* = \frac{qm^* k_B^2}{2\pi^2\hbar^3} \approx 120\,\mathrm{A/cm^2\,K^2} \quad \text{for } m^* = m_0. \tag{S33}$$

This similarity suggests that the Tsu–Esaki model extends the classical Richardson emission by incorporating sub-barrier tunneling through a transmission coefficient $T(E) \ll 1$. While the Richardson model assumes perfect transmission $T(E) = 1$ above the barrier (and $T = 0$ below it), the Tsu–Esaki model explicitly accounts for quantum tunneling across the energy barrier. The gate voltage dependence enters through both the supply function and the barrier height, leading to an exponential modulation of the tunneling current.

In the case of the insulator–$MoS_2$ stacks examined in this work, the gate metal work function is taken as 5.2 eV and the $MoS_2$ electron affinity as 4.3 eV, resulting in a barrier height of $\phi_B = 0.9$ eV (Eq. S30). To evaluate the tunneling current according to IRDS guidelines for 2037 nodes, a gate voltage of 0.6 V was assumed which yields

$$\boxed{J_0 \approx 100\ \mathrm{A/cm^2} \quad \text{with} \quad \phi_B = 0.9\ \mathrm{eV}, \quad V_G = 0.6\ \mathrm{V}.} \tag{S34}$$

**Remark:** Although this simplified model provides a closed-form prefactor $J_0$ and captures the leading scaling trends relevant to minimum EOT estimation (Section S5), it does not replace numerical calculations and is used solely for first-order analytical insight. All results shown in the figures are obtained from the rigorous numerical model.

## S4 Two-band model

When electrons tunnel through a wide-gap insulator, their wavefunctions decay exponentially in the classically forbidden energy range. This behavior is described by the complex band structure, which extends conventional band theory to complex wavevectors. Inside the gap, propagating Bloch states become evanescent states characterized by an imaginary wavevector component. The decay rate is quantified by the inverse decay length, defined as $\beta(E) = 2\,\mathrm{Im}\{k(E)\}$.

A widely used approximation is the two-band Franz model (*58*), which assumes symmetric parabolic conduction and valence bands. It expresses $\beta(E)$ as:

$$\beta(E) = \frac{2}{\hbar}\sqrt{2m^* \frac{(E - E_\mathrm{v})(E_\mathrm{c} - E)}{E_\mathrm{g}}}, \tag{S35}$$

where $m^*$ is the tunneling effective mass, $E_\mathrm{v}$ and $E_\mathrm{c}$ are the valence and conduction band edges, and $E_\mathrm{g} = E_\mathrm{c} - E_\mathrm{v}$ is the band gap. In terms of an energy offset $\Delta E$ from a band edge, Eq. S35 becomes:

$$\boxed{\beta(\Delta E) = \frac{2}{\hbar}\sqrt{2m^* \Delta E \left(1 - \frac{\Delta E}{E_\mathrm{g}}\right)},} \tag{S36}$$

with $\Delta E = E_\mathrm{c} - E$ for electrons and $\Delta E = E - E_\mathrm{v}$ for holes. This form clearly shows that the decay is minimal near the bandedges and has a maximum at midgap.

To account for different effective masses in the conduction ($m_\mathrm{c}$) and valence ($m_\mathrm{v}$) bands, the model can be generalized into an elliptic two-band approximation (*59*), yielding:

$$\beta(\Delta E) = \begin{cases} \frac{2}{\hbar}\sqrt{2m_\mathrm{v}\,\Delta E \left(1 - \frac{\Delta E}{2E_{\mathrm{q,v}}}\right)}, & \text{holes: } 0 \le \Delta E = E - E_\mathrm{v} \le E_{\mathrm{q,v}}, \\ \frac{2}{\hbar}\sqrt{2m_\mathrm{c}\,\Delta E \left(1 - \frac{\Delta E}{2E_{\mathrm{q,c}}}\right)}, & \text{electrons: } 0 \le \Delta E = E_\mathrm{c} - E \le E_{\mathrm{q,c}}, \end{cases} \tag{S37}$$

where the transition energies are:

$$E_{\mathrm{q,v}} = \frac{m_\mathrm{c}}{m_\mathrm{c} + m_\mathrm{v}}\,E_\mathrm{g}, \quad E_{\mathrm{q,c}} = \frac{m_\mathrm{v}}{m_\mathrm{c} + m_\mathrm{v}}\,E_\mathrm{g}.$$

The decay reaches its maximum at the branch-point energy $E_\mathrm{q} = E_\mathrm{v} + E_{\mathrm{q,v}} = E_\mathrm{c} - E_{\mathrm{q,c}}$.

For electron tunneling from $SiO_2$ into Si, the relevant barrier is the conduction band offset $\Delta E_{\mathrm{SiO_2}} \approx 3.1$ eV (*60, 61*). Using $m_\mathrm{c} \approx 0.42\,m_0$, $m_\mathrm{v} \approx 0.33\,m_0$, and $E_\mathrm{g} = 8.9$ eV for $SiO_2$, the resulting inverse decay length from Eq. S37 is:

$$\boxed{\beta_{\mathrm{SiO_2}} \simeq 0.91\ \text{Å}^{-1}.} \tag{S38}$$

## S5 Insulator figure of merit and minimum achievable EOT

As described in Section S3, the gate leakage current in a metal–insulator–semiconductor structure is strongly influenced by the transmission probability $T(E)$ through the insulator, which can be approximated using the WKB method:

$$T(E) \approx \exp\Big[-\int_{z_1}^{z_2} \beta(z,E)\,\mathrm{d}z\Big], \qquad \text{with} \quad \beta(z,E) = \frac{2}{\hbar}\sqrt{2\,m^*\,\Delta E(z,E)} \tag{S39}$$

where $\beta(z,E)$ is the local inverse decay length and $\Delta E(z,E)$ the energy barrier between the carrier energy $E$ and the relevant band edge of the insulator: $\Delta E = E_{\mathrm{c,ins}}(z) - E$ for electrons and $\Delta E = E - E_{\mathrm{v,ins}}(z)$ for holes. The analysis that follows focuses on electrons; analogous relations apply for holes. In the flat-band limit, where the band edges are spatially uniform, $E_{\mathrm{c,ins}}(z) \simeq E_{\mathrm{c,ins}}$, the integral simplifies to

$$T(E) \approx \exp(-\beta(E)\,t_{\mathrm{ins}})\,, \tag{S40}$$

with $\beta(E) = 2\sqrt{2\,m^*\,\Delta E}/\hbar$, $\Delta E = E_{\mathrm{c,ins}} - E$ for electron tunnelling, and $t_{\mathrm{ins}}$ denotes the physical thickness of the dielectric. Because the Fermi–Dirac tail decays exponentially, the dominant contribution to the integral stems from electrons whose longitudinal energy lies close to the band edge of the semiconductor channel. The transmission is therefore evaluated at the channel conduction-band edge $E_{\mathrm{c,ch}}$. As illustrated in Fig. S8B, the relevant barrier height is the conduction-band offset,

$$\Delta E = E_{\mathrm{c,ins}} - E_{\mathrm{c,ch}} = \chi_{\mathrm{ch}} - \chi_{\mathrm{ins}},$$

where $\chi_{\mathrm{ch}}$ and $\chi_{\mathrm{ins}}$ denote, respectively, the electron affinities of the channel and of the insulator. (For hole tunnelling the analogous quantity is $\Delta E = E_{\mathrm{v,ch}} - E_{\mathrm{v,ins}}$.) This band-offset $\Delta E$ is the height that enters the inverse decay length $\beta(E)$. Using the simplified Tsu–Esaki model in Eq. S31, one can approximately evaluate the tunneling current as:

$$J \approx J_0\,T(E_{\mathrm{c,ch}}) = J_0 \exp(-\beta\,t_{\mathrm{ins}})\,, \tag{S41}$$

where $J_0$ includes prefactors and the supply term and the inverse of the decay length is given by

$$\beta = \frac{2}{\hbar}\sqrt{2m^*\Delta E} = \frac{2}{\hbar}\sqrt{2m^*\left(E_{\mathrm{c,ins}} - E_{\mathrm{c,ch}}\right)} \tag{S42}$$

To provide a more accurate reference for benchmarking tunneling suppression, one can utilize the decay constant using the complex band structure formalism described in Eq. S36 or Eq. S37.

The physical thickness is related to electrostatic scaling by expressing it in terms of the required EOT of a target node:

$$t_{\mathrm{ins}} = \frac{\kappa}{\kappa_{\mathrm{SiO_2}}}\mathrm{EOT}, \tag{S43}$$

where $\kappa$ is the permittivity of the insulator. Substituting the above into the current relation we obtain

$$J = J_0 \exp\left(-\beta\frac{\kappa}{\kappa_{\mathrm{SiO_2}}}\mathrm{EOT}\right). \tag{S44}$$

It is convenient to normalise every material to thermally-grown $SiO_2$ by factoring out its WKB decay parameter expressed in Eq. S38. Thereby, the dimensionless figure of merit (FoM) is introduced using this reference:

$$\boxed{\mathrm{FoM} \;=\; \frac{\kappa}{\kappa_{\mathrm{SiO_2}}}\,\frac{\beta}{\beta_{\mathrm{SiO_2}}} \;=\; \frac{\kappa}{\kappa_{\mathrm{SiO_2}}}\sqrt{\frac{m^*\,\Delta E}{m^*_{\mathrm{SiO_2}}\,\Delta E_{\mathrm{SiO_2}}}},} \tag{S45}$$

so that the leakage current can be written in a compact form as

$$\boxed{J \;=\; J_0\,\exp\!\left[-\,\beta_{\mathrm{SiO_2}}\,\mathrm{FoM}\;\mathrm{EOT}\right].} \tag{S46}$$

By construction, the FoM is normalized to unity for $SiO_2$. Values larger (smaller) than 1 indicate superior (inferior) suppression of gate tunneling per unit EOT. This is because the FoM encapsulates both electrostatic enhancement, via the permittivity $\kappa$, and quantum tunneling suppression, via the inverse decay length $\beta$. Accordingly, the dimensionless FoM introduced in Eq. S45 can also be refined using the two-band model as shown in Section S4.

Physically, the FoM quantifies how effectively a given insulator suppresses the tunneling current for each unit increase in EOT. It is proportional to the product of the permittivity $\kappa$ and the square root of the product of the tunneling effective mass and the conduction band offset between the insulator and the channel material (e.g., $MoS_2$). This reflects the decay rate of the electron wavefunction across the dielectric. A larger FoM implies stronger suppression of gate leakage for a given EOT, thereby enabling more aggressive scaling of gate control while maintaining acceptable leakage limits. Rearranging Eq. S45 yields the minimum achievable EOT under a fixed leakage current target $J = J_{\mathrm{target}}$, where the IRDS roadmap specifies $J_{\mathrm{target}} = 1.5 \times 10^{-2}$ A/cm$^2$ for low-power applications (*3*) as:

$$\boxed{\mathrm{EOT}_{\mathrm{min}} = \frac{1}{\beta_{\mathrm{SiO_2}}\mathrm{FoM}}\ln\left(\frac{J_0}{J_{\mathrm{target}}}\right).} \tag{S47}$$

**Dead layer effects**: Incorporation of dead layer effects is achieved via the effective permittivity model outlined in Section S2. By inserting the effective permittivity from Eq. S22 into the EOT relation in Eq. S43, the resulting expression is:

$$\mathrm{EOT} = \left( \frac{t}{\kappa_{\mathrm{bulk}}} + D \right) \kappa_{\mathrm{SiO_2}}. \tag{S48}$$

Solving for the physical thickness $t$ yields:

$$t = \frac{\kappa_{\mathrm{bulk}}}{\kappa_{\mathrm{SiO_2}}} \left( \mathrm{EOT} - D\kappa_{\mathrm{SiO_2}} \right). \tag{S49}$$

Upon substitution of this thickness into Eq. S41, the current is given by:

$$J = J_0 \exp \left[ -\beta \frac{\kappa_{\mathrm{bulk}}}{\kappa_{\mathrm{SiO_2}}} \left( \mathrm{EOT} - D\kappa_{\mathrm{SiO_2}} \right) \right]. \tag{S50}$$

This can be rearranged as:

$$J = J_0 \exp \left[ -\beta_{\mathrm{SiO_2}} \mathrm{FoM}_{\mathrm{bulk}} \left( \mathrm{EOT} - D\kappa_{\mathrm{SiO_2}} \right) \right] \tag{S51}$$

where the bulk FoM, in the absence of a dead layer, is defined as:

$$\mathrm{FoM}_{\mathrm{bulk}} = \frac{\kappa_{\mathrm{bulk}}}{\kappa_{\mathrm{SiO_2}}} \frac{\beta}{\beta_{\mathrm{SiO_2}}}. \tag{S52}$$

The minimum EOT corresponding to a fixed target current $J = J_{\mathrm{target}}$ is obtained as:

$$\boxed{\mathrm{EOT}_{\mathrm{total}}^{\mathrm{min}} = \mathrm{EOT}_{\mathrm{bulk}}^{\mathrm{min}} + \mathrm{EOT}_{\mathrm{dead}} = \frac{1}{\beta_{\mathrm{SiO_2}} \mathrm{FoM}_{\mathrm{bulk}}} \ln \left( \frac{J_0}{J_{\mathrm{target}}} \right) + D\kappa_{\mathrm{SiO_2}}.} \tag{S53}$$

This final expression separates two key contributions: an ideal minimum EOT term determined by bulk material properties, and an additive penalty from dead layer effects. For example, although bulk $SrTiO_3$ (STO) has a permittivity exceeding 270 at room temperature, dead layers greatly reduce the effective response in ultrathin films, limiting the benefit of STO's high-$\kappa$ permittivity. Using reported STO parameters from (*18*), with a minimal dead layer parameter of $D = 1.6$ Å, the increase in minimum achievable EOT is:

$$\mathrm{EOT}_{\mathrm{dead}}^{\mathrm{STO}} = D \times \kappa_{\mathrm{SiO_2}} = 1.61 \text{ Å} \times 3.9 \approx 6.3 \text{ Å}. \tag{S54}$$

This offset imposes a hard limit on gate scaling with ultra high-$\kappa$ dielectrics like STO.

**Electronic polarizability in dead layer models**: In the previous section, it was assumed that the permittivity scales nearly inversely with thickness. This effect is expected to originate primarily from the ionic contribution to permittivity, which is degraded by interfacial effects. By contrast, the electronic permittivity can be reasonably assumed to remain largely unaffected by thickness variation. Considering this behavior, the minimum achievable EOT in the dead layer model must be revised. It is assumed here that $\kappa^{\mathrm{el}}$ remains nearly constant, while the ionic component $\kappa^{\mathrm{ion}}$ follows a thickness-dependent dead layer model. The total effective permittivity is expressed as the sum of the electronic and ionic contributions: $\kappa_{\mathrm{eff}} = \kappa^{\mathrm{el}} + \kappa^{\mathrm{ion}}$. According to this modified dead layer model, only the ionic permittivity is modified as follows:

$$\frac{t}{\kappa^{\mathrm{ion}}} = \frac{t}{\kappa^{\mathrm{ion}}_{\mathrm{bulk}}} + D \quad \Rightarrow \quad \kappa^{\mathrm{ion}} = \frac{\kappa^{\mathrm{ion}}_{\mathrm{bulk}}}{1 + D\kappa^{\mathrm{ion}}_{\mathrm{bulk}}/t}. \tag{S55}$$

For simplicity, we define the bulk permittivity $\kappa_{\mathrm{bulk}} = \kappa^{\mathrm{el}} + \kappa^{\mathrm{ion}}_{\mathrm{bulk}}$ and $\eta = \kappa^{\mathrm{el}}/\kappa_{\mathrm{bulk}}$, with $1 - \eta = \kappa^{\mathrm{ion}}_{\mathrm{bulk}}/\kappa_{\mathrm{bulk}}$. With these definitions, the effective permittivity becomes:

$$\kappa_{\mathrm{eff}} = \kappa_{\mathrm{bulk}} \left( \eta + (1-\eta) \frac{1}{1 + D(1-\eta)\kappa_{\mathrm{bulk}}/t} \right). \tag{S56}$$

The corresponding physical thickness associated with $\mathrm{EOT}_{\mathrm{min}}$ is given by:

$$t_{\mathrm{min}} = \frac{\ln(J_0/J_{\mathrm{target}})}{\beta}, \quad \text{where} \quad \beta = \frac{\mathrm{FoM}^{\mathrm{bulk}}\beta_{\mathrm{SiO_2}}}{\kappa_{\mathrm{bulk}}}. \tag{S57}$$

Substituting $t_{\mathrm{min}}$ and $\beta$ yields the minimum achievable EOT:

$$\mathrm{EOT}_{\mathrm{min}} = t_{\mathrm{min}}\kappa_{\mathrm{SiO_2}}/\kappa_{\mathrm{eff}} = \frac{\ln(J_0/J_{\mathrm{target}})}{\beta}\kappa_{\mathrm{SiO_2}} \left[ \kappa_{\mathrm{bulk}} \left( \eta + \frac{1-\eta}{1 + (1-\eta)D\kappa_{\mathrm{bulk}}/t} \right) \right]^{-1}. \tag{S58}$$

After simplification, and using the definition of the bulk figure of merit,

$$\mathrm{FoM}^{\mathrm{bulk}} = \frac{\kappa_{\mathrm{bulk}}\beta}{\kappa_{\mathrm{SiO_2}}\beta_{\mathrm{SiO_2}}}, \qquad \mathrm{EOT}^{\mathrm{bulk}}_{\mathrm{min}} = \frac{\ln(J_0/J_{\mathrm{target}})}{\mathrm{FoM}^{\mathrm{bulk}}\beta_{\mathrm{SiO_2}}}, \tag{S59}$$

the final compact expression becomes:

$$\mathrm{EOT}_{\mathrm{min}} = \mathrm{EOT}^{\mathrm{bulk}}_{\mathrm{min}} \left[ \eta + \frac{1-\eta}{1 + (1-\eta)D\kappa_{\mathrm{SiO_2}}/\mathrm{EOT}^{\mathrm{bulk}}_{\mathrm{min}}} \right]^{-1}. \tag{S60}$$

This general result reduces to the pure dead layer model when $\kappa^{\mathrm{el}} \to 0 \Rightarrow \eta \to 0$, and recovers the bulk permittivity model in the limit $D \to 0$.

**vdW gap effects**: To assess the impact of the vdW gap on leakage and scaling, the gate stack is modeled as two sequential tunneling barriers: (i) an insulator of thickness $t_{\text{ins}}$ and permittivity $\kappa_{\text{ins}}$, and (ii) a vdW gap of thickness $t_{\text{vdW}}$ and permittivity $\kappa_{\text{vdW}}$. Within the WKB approximation and assuming phase-coherent transport across the ultrathin insulator and vdW gap, the total tunneling current density through the stack can be approximated by the product of the transmission probabilities through the individual regions. Accordingly, the total current density is given by: $J \approx J_0 \exp\left(-\beta_{\text{ins}} t_{\text{ins}} - \beta_{\text{vdW}} t_{\text{vdW}}\right)$, where $\beta_{\text{ins}}$ and $\beta_{\text{vdW}}$ are the inverse decay lengths in the insulator and the vdW gap, respectively. Electrostatic scaling is captured by relating the physical thicknesses to their EOT contributions via:

$$t_{\text{ins}} = \frac{\kappa_{\text{SiO}_2}}{\kappa_{\text{ins}}} \text{EOT}_{\text{ins}}, \qquad t_{\text{vdW}} = \frac{\kappa_{\text{SiO}_2}}{\kappa_{\text{vdW}}} \text{EOT}_{\text{vdW}}. \tag{S61}$$

Substituting into the tunneling expression and using the FoM from Eq. S45 yields:

$$J = J_0 \exp\left[-\beta_{\text{SiO}_2} \left(\text{FoM}_{\text{ins}} \text{EOT}_{\text{ins}} + \text{FoM}_{\text{vdW}} \text{EOT}_{\text{vdW}}\right)\right], \tag{S62}$$

where the following dimensionless FoMs were defined:

$$\text{FoM}_{\text{ins}} = \frac{\kappa_{\text{ins}}}{\kappa_{\text{SiO}_2}} \sqrt{\frac{m^*_{\text{ins}} \Delta E_{\text{ins}}}{m^*_{\text{SiO}_2} \Delta E_{\text{SiO}_2}}}, \qquad \text{FoM}_{\text{vdW}} = \frac{\kappa_{\text{vdW}}}{\kappa_{\text{SiO}_2}} \sqrt{\frac{m^*_{\text{vdW}} \Delta E_{\text{vdW}}}{m^*_{\text{SiO}_2} \Delta E_{\text{SiO}_2}}}. \tag{S63}$$

For the insulator, the conduction band offset is given by $\Delta E_{\text{ins}} = \chi_{\text{MoS}_2} - \chi_{\text{ins}}$. Tunneling through the vdW gap is modeled as vacuum tunneling, consistent with DFT results (Fig. 4B), using a barrier height $\Delta E_{\text{vdW}} = \chi_{\text{MoS}_2}$. Lacking lattice and bonding, the vdW gap imposes vacuum-like conditions, justifying the use of $m^*_{\text{vdW}} = m_0$. To determine the minimum achievable EOT under a fixed gate leakage target $J = J_{\text{target}}$, the tunneling exponent must satisfy:

$$\beta_{\text{SiO}_2} \left(\text{FoM}_{\text{ins}} \text{EOT}_{\text{ins}} + \text{FoM}_{\text{vdW}} \text{EOT}_{\text{vdW}}\right) = \ln(J_0/J_{\text{target}}). \tag{S64}$$

Solving for $\text{EOT}_{\text{ins}}$ gives:

$$\text{EOT}_{\text{ins}} = \frac{\ln(J_0/J_{\text{target}}) - \beta_{\text{SiO}_2} \text{FoM}_{\text{vdW}} \text{EOT}_{\text{vdW}}}{\beta_{\text{SiO}_2} \text{FoM}_{\text{ins}}}, \tag{S65}$$

where the first term, $\text{EOT}^{\text{min}}_{\text{ins}} = \ln(J_0/J_{\text{target}})/(\beta_{\text{SiO}_2} \text{FoM}_{\text{ins}})$, represents the minimum achievable EOT in the absence of a vdW gap; the total minimum achievable EOT is therefore given by:

$$\boxed{\text{EOT}^{\min}_{\text{total}} = \text{EOT}^{\min}_{\text{ins}} + \text{EOT}_{\text{vdW}} = \frac{\ln(J_0/J_{\text{target}})}{\beta_{\text{SiO}_2}\text{FoM}_{\text{ins}}} + \left(1 - \frac{\text{FoM}_{\text{vdW}}}{\text{FoM}_{\text{ins}}}\right)\text{EOT}_{\text{vdW}}.} \tag{S66}$$

The second term represents the modification introduced by the vdW layer. It contributes negatively–i.e., beneficially–only when: $\text{FoM}_{\text{ins}} < \text{FoM}_{\text{vdW}}$. Substituting the FoM definitions into this inequality and using $m^*_{\text{ins}} = m_r m_0$, $\Delta E_{\text{vdW}} = \chi_{\text{MoS}_2}$, and $\Delta E_{\text{ins}} = \chi_{\text{MoS}_2} - \chi_{\text{ins}}$, the condition simplifies to:

$$m_r\left(1 - \frac{\chi_{\text{ins}}}{\chi_{\text{MoS}_2}}\right) < \left(\frac{\kappa_{\text{vdW}}}{\kappa_{\text{ins}}}\right)^2. \tag{S67}$$

This inequality defines the regime where the inclusion of a vdW gap improves the leakage–EOT tradeoff by offering more effective tunneling suppression per unit electrostatic thickness than the insulator alone. To illustrate this condition, Fig. S9A presents the boundary in the $(\kappa_{\text{ins}}, \chi_{\text{ins}})$ space beyond which a vdW gap reduces the minimum achievable EOT. A clear dependence of the normalized EOT modification, $\Delta\text{EOT}/\text{EOT}_{\text{vdW}}$, on the insulator permittivity for various representative parameters is illustrated in Fig. S9B.

**vdW gap average EOT and FoM**: For all studied insulator–$MoS_2$ stacks, an analytical model of the form $\kappa^{-1}_{\text{vdW}} = 1 - c/t_{\text{vdW}}$, derived in Eq. S6, was fitted to the extracted permittivities. For each insulator–$MoS_2$ interface, a best-fit parameter $c$ was obtained individually. The average value across all systems was found to be $\bar{c} = 0.72$ Å, as summarized in Table S2. Based on this model and the fitted $\bar{c}$, the average permittivity and EOT of the vdW gap were determined as $\bar{\kappa}_{\text{vdW}} \approx 2$ and $\overline{\text{EOT}}_{\text{vdW}} \approx 2.7$ Å, respectively.

For electron tunneling through the vdW gap, a barrier height equal to the electron affinity of $MoS_2$, $\chi_{\text{MoS}_2} = 4.30$ eV, and a shape factor of $\alpha = 0.8$ were assumed. The corresponding tunneling decay constant is:

$$\beta_{\text{vdW}} = \frac{2\alpha}{\hbar}\sqrt{2\, m_0\, \chi_{\text{MoS}_2}} = 1.7\ \text{Å}^{-1}. \tag{S68}$$

From these values, the FoM for the vdW gap is calculated as:

$$\boxed{\overline{\text{FoM}}_{\text{vdW}} = \frac{\bar{\kappa}_{\text{vdW}} \times \beta_{\text{vdW}}}{\kappa_{\text{SiO}_2} \times \beta_{\text{SiO}_2}} = \frac{2.03 \times 1.7}{3.9 \times 0.91} \approx 1,} \tag{S69}$$

where $\beta_{\text{SiO}_2} = 0.91\ \text{Å}^{-1}$ was obtained from Eq. S38.

Finally, the minimum achievable total EOT, including the vdW gap, can be expressed as:

$$\text{EOT}^{\min}_{\text{total}} = \frac{8.79}{0.91 \times \text{FoM}_{\text{ins}}} + \left(1 - \frac{1}{\text{FoM}_{\text{ins}}}\right)2.73 = \frac{7\ \text{Å}}{\text{FoM}_{\text{ins}}} + 2.73\ \text{Å}. \tag{S70}$$

## S6 Quantum contact resistance with a vdW gap

In contacts between a 3D metal and a low-dimensional material (such as a 2D semiconductor), an intrinsic quantum contact resistance arises from the finite number of available conduction modes in the 2D channel that couple with an ideal contact. A recent comprehensive derivation for 2D semiconductors gives the width-normalized form (*19*):

$$R_{\text{cq}}W \;=\; \frac{h}{2e^2}\frac{\sqrt{\pi/2}}{\sqrt{g_v n_{\text{2D}}}} \qquad \text{(parabolic band, } T = 1\text{)}, \tag{S71}$$

where $g_v$ is the valley degeneracy and $n_{\text{2D}}$ is the 2D carrier density; cf. Eq. (5) and related discussion in Ref. (*19*). The above relation corresponds to the total resistance of a ballistic device with two ideal contacts, i.e., $R_{\text{tot}} \to R_{\text{cq}} = 2R_c$ as the channel becomes dissipation-free.

In realistic 2D metal-semiconductor junctions, however, an atomically thin vdW gap can form at the interface. Its principal effect in the ground-state Landauer picture is to introduce an energy-dependent transmission $T(E) \le 1$, so the conductance becomes

$$G \;=\; \frac{2e^2}{h} M_{\text{eff}} \langle T \rangle \quad\Longrightarrow\quad R_{\text{cq}}W \;=\; \frac{h}{2e^2}\frac{W}{M_{\text{eff}}\langle T \rangle}, \tag{S72}$$

with $M_{\text{eff}} = \frac{Wg_v}{\pi} k_F$ for a 2D parabolic band at zero temperature and $\langle T \rangle$ is the Fermi-window–averaged transmission $\langle T \rangle = \int dE\, T(E)(-\partial f/\partial E)$, which equals $T(E_F)$ at $T = 0$. Using $k_F = \sqrt{2\pi n_{\text{2D}}/g_v}$, Eq. S72 reduces to Eq. S71 when $\langle T \rangle = 1$.

To model a vdW gap of thickness $t_{\text{vdW}}$ and barrier height $\Delta E$ with free-electron mass $m_0$, we adopt the usual WKB transmission for a rectangular barrier:

$$T_{\text{vdW}} \;\equiv\; \exp\!\left[-\frac{2\alpha\, t_{\text{vdW}}}{\hbar}\sqrt{2m_0\Delta E}\right], \tag{S73}$$

where $\alpha$ is a unitless shape factor for the vdW barrier and $\hbar$ is the reduced Planck constant. Substituting Eq. S73 and the expression for $M_{\text{eff}}$ into Eq. S72 yields the vdW-corrected quantum contact resistance:

$$R_{\text{cq}}^{(\text{vdW})}W \;=\; \frac{h}{2e^2}\frac{\sqrt{\pi/2}}{\sqrt{g_v n_{\text{2D}}}}\frac{1}{T_{\text{vdW}}} \;=\; \left(R_{\text{cq}}W\right)_{T=1}\,\exp\!\left[\frac{2\alpha\, t_{\text{vdW}}}{\hbar}\sqrt{2m_0\Delta E}\right]. \tag{S74}$$

Equation Eq. S74 reduces to the ideal case of Ref. (*19*) when $t_{\text{vdW}} \to 0$, i.e., $T_{\text{vdW}} \to 1$. In other words, the vdW gap suppresses the interfacial mode transmission by the factor $T_{\text{vdW}}$, so the ballistic conductance is reduced by $T_{\text{vdW}}$ and the quantum contact resistance increases by $1/T_{\text{vdW}}$.

## S7 Validation of the vdW analytical tunneling model by first-principles transport

We benchmark the compact vdW tunneling model against first-principles references—*ab initio* NEGF–DFT for a representative Au–$MoS_2$ junction. Computational details are provided in Materials and Methods. The left electrode is fcc Au and the right electrode is $MoS_2$; the in-plane Bravais vectors are identical for the electrodes and the central region to ensure commensurability. Because a vdW interface is intrinsically characterized by a finite interfacial separation, it cannot be continuously transformed into a gap-free contact without entering a covalently bonded regime. We therefore validate the analytical model by systematically increasing the vdW separation and monitoring the corresponding evolution of the transmission and planar-averaged potential, which isolates the tunneling contribution and provides a robust, quantitative benchmark.

Concretely, we enlarged the Au–$MoS_2$ spacing relative to the relaxed interface, increasing the distance between the adjacent interfacial atomic planes from $d_{\text{Au-S}} = 4.1$ Å to 8.1 Å in 1 Å increments, and for each geometry computed the planar-averaged electrostatic potential and the transmission. For numerically stable vacuum-like barriers at larger separations, basis-only ghost atoms (no ionic potential) were inserted in the gap to preserve basis completeness.

The planar-averaged electrostatic potential $\overline{V}(z)$ across the Au–$MoS_2$ interface is shown in Fig. S10A. Because the potential maximum in the gap approaches the vacuum level, the injection barrier into the $MoS_2$ conduction band is well approximated by the $MoS_2$ electron affinity, see Fig. S10B. The corresponding near–conduction-band-edge transmissions in Fig. S10C decay exponentially with increasing separation. This behavior is captured by a WKB model with a vacuum-like effective mass and a rectangular barrier ($\alpha = 1$), where the change in the effective tunneling width is identified as $\Delta t_{\text{vdW}} \equiv \Delta d_{\text{Au-S}}$, such that

$$\frac{T_{\text{vdW}}\big(d_{\text{Au-S}} + \Delta t_{\text{vdW}}\big)}{T_{\text{vdW}}(d_{\text{Au-S}})} = \exp\big[-\beta\,\Delta t_{\text{vdW}}\big], \qquad m^* \approx m_0,\ \Delta E \approx 4.3\,\text{eV},\ \beta = 2\frac{\sqrt{2m^*\Delta E}}{\hbar} \approx 2.12\,\text{Å}^{-1}.$$

A 1 Å increase yields $T_{\text{vdW}}(d_{\text{Au-S}}{+}1\,\text{Å})/T_{\text{vdW}}(d_{\text{Au-S}}) = \exp(-2.12) \approx 0.12$ (i.e., a $\sim 10\times$ reduction), in excellent agreement with the NEGF results. From $d_{\text{Au-S}} = 5.1$ to 8.1 Å, each additional 1 Å increase produces roughly a tenfold drop in transmission, consistent with our calculations. The stronger reduction at the smallest separation ($d_{\text{Au-S}} = 4.1$ Å) is explained by a slightly trapezoidal barrier profile; in this regime, both the increased separation and the larger effective shape factor $\alpha$ contribute to the enhanced suppression of the transmission.

## S8 $MoS_2$ channel contribution to capacitance-equivalent thickness

IRDS scaling projections for beyond-2030 technology nodes target total gate-stack capacitance-equivalent thickness (CET) values CET < 9 Å. In bulk Si MOS systems, inversion-layer centroid displacement and quantum-confinement effects introduce an effective electrostatic thickness on the order of ~ 4 Å, which adds in series with the gate insulator. Consequently, achieving CET < 9 Å would require insulator equivalent oxide thickness (EOT) values $EOT_{ins} < 5$ Å in advanced Si technologies.

For atomically thin 2D semiconductors, the corresponding channel contribution requires careful quantification, particularly in the presence of a finite vdW gap, where electrostatic partitioning can substantially modify the effective gate coupling. Bennett and Pop identified two primary mechanisms limiting monolayer $MoS_2$ electrostatics: (i) finite density of states (quantum capacitance) and (ii) charge-centroid capacitance arising from nonuniform carrier distributions across the channel thickness (*62*). The present work extends this framework by incorporating an additional vdW gap capacitance term, which is shown in the main manuscript to impose a substantial scaling penalty.

The purpose of this section is to quantitatively establish the intrinsic $MoS_2$ channel contribution to CET in an insulator–$MoS_2$ stack with a vdW gap, which is required to determine the ultimate insulator EOT limits for 2D transistor scaling. The terminal gate capacitance is defined as

$$C_G \equiv \frac{dQ}{dV_G} = q\,\frac{\partial n_{2D}}{\partial V_G}, \tag{S75}$$

where $n_{2D}$ is the channel sheet carrier density and $Q \equiv q\,n_{2D}$ is the sheet charge density. Throughout this section, all capacitances are normalized per unit area. For a finite insulator thickness, the gate voltage partitions between the insulator and the 2D channel as

$$V_G = \psi_{MoS_2} + \frac{Q}{C_{ins}}, \tag{S76}$$

where $\psi_{MoS_2}$ is the electrostatic potential in the $MoS_2$ monolayer and $C_{ins}$ is the insulator capacitance per unit area. The intrinsic channel capacitance is defined as

$$C_{MoS_2}(\psi_{MoS_2}) \equiv \frac{dQ}{d\psi_{MoS_2}} = q\,\frac{dn_{2D}}{d\psi_{MoS_2}}. \tag{S77}$$

Differentiating Eq. S76 with respect to $V_G$ yields

$$1 = \frac{d\psi_{MoS_2}}{dV_G} + \frac{1}{C_{ins}}\frac{dQ}{dV_G}, \tag{S78}$$

which, using $\mathrm{d}Q/\mathrm{d}V_\mathrm{G} = C_\mathrm{G}$ and $\mathrm{d}Q/\mathrm{d}\psi_{\mathrm{MoS_2}} = C_{\mathrm{MoS_2}}$, gives the series relation

$$\frac{1}{C_\mathrm{G}} = \frac{1}{C_\mathrm{ins}} + \frac{1}{C_{\mathrm{MoS_2}}}. \tag{S79}$$

If a vdW gap is present, it is treated as an additional series element so that

$$C_\mathrm{G}^{-1} = C_\mathrm{ins}^{-1} + C_\mathrm{vdW}^{-1} + C_{\mathrm{MoS_2}}^{-1}. \tag{S80}$$

The insulator and vdW contributions are expressed as

$$\mathrm{EOT_{ins}} \equiv \frac{\kappa_{\mathrm{SiO_2}}}{C_\mathrm{ins}}, \quad \mathrm{EOT_{vdW}} \equiv \frac{\kappa_{\mathrm{SiO_2}}}{C_\mathrm{vdW}}, \tag{S81}$$

and the $MoS_2$ channel contribution to CET is

$$\mathrm{CET}_{\mathrm{MoS_2}} \equiv \frac{\kappa_{\mathrm{SiO_2}}}{C_{\mathrm{MoS_2}}}. \tag{S82}$$

The total gate-stack CET is therefore

$$\boxed{\mathrm{CET_{total}} = \mathrm{EOT_{ins}} + \mathrm{EOT_{vdW}} + \mathrm{CET}_{\mathrm{MoS_2}}.} \tag{S83}$$

The intrinsic channel capacitance reflects both the finite density of states (DOS) and bias-dependent charge-centroid effects captured self-consistently from DFT-derived local density of states (LDOS) (*62*). As a useful upper bound, the 2D quantum capacitance for a parabolic band is

$$C_Q = q^2 g_v g_s \frac{m^*}{\pi\hbar^2}, \tag{S84}$$

where $m^*$ is the effective mass and $g_v$ and $g_s$ are the valley and spin degeneracies. Using $m^* \approx 0.45m_0$, $g_v = 2$ (for K valley), and $g_s = 2$ yields $C_Q \approx 60$ µF/cm$^2$. In practice, allowing a nonuniform potential across the monolayer thickness produces a bias-dependent charge centroid that reduces $C_{\mathrm{MoS_2}}$ below this ideal limit, even when the channel is degenerate.

The midgap of $MoS_2$ is used as the energy reference. A convenient surface-potential threshold reference is defined as (*62*)

$$\psi_\mathrm{th} \equiv \frac{E_G}{2q}, \tag{S85}$$

corresponding to the midgap-to-conduction-band-edge energy separation. Under electrostatic bias, the surface potential shifts such that the Fermi level aligns with the conduction-band edge $E_\mathrm{c}$; this condition is used here as a reference for the onset of degenerate accumulation (inset of Fig. S11B).

In general, the terminal threshold voltage can be written as

$$V_{th} = \psi_{th} + \Delta V_{FB} + \frac{Q_{th}}{C_{ins}}, \quad \text{(S86)}$$

where $\Delta V_{FB}$ denotes the effective flat-band shift due to gate–channel work-function difference and fixed oxide/interface charges (set to zero for clarity, $\Delta V_{FB} = 0$), and $Q_{th} \equiv qn_{2D,th}$ is the *magnitude* of the threshold sheet charge density.

Channel quantities are reported versus $\psi_{MoS_2} - \psi_{th}$, and terminal quantities versus $V_G - V_{th}$, with $V_{th} \approx \psi_{th}$ as discussed below. At $\psi_{MoS_2} = \psi_{th}$, the extracted sheet carrier density is $n_{2D,th} \approx 2 \times 10^{12}\ \text{cm}^{-2}$ (inset of Fig. S11A), which is adopted as a practical threshold-like reference for the onset of accumulation. For the representative oxide scaling case used in Fig. S11, we use EOT = 5 Å, giving

$$C_{ins} = \frac{\kappa_{SiO_2}}{\text{EOT}} \approx 6.9\ \mu\text{F/cm}^2. \quad \text{(S87)}$$

The gate-voltage partition term at the threshold reference density is then

$$\Delta V_{ins,th} \equiv \frac{Q_{th}}{C_{ins}} = \frac{q\, n_{2D,th}}{C_{ins}} \approx \frac{(1.6 \times 10^{-19}\ \text{C})(2 \times 10^{12}\ \text{cm}^{-2})}{6.9 \times 10^{-6}\ \text{F/cm}^2} \approx 4.6 \times 10^{-2}\ \text{V} \approx 50\ \text{mV}, \quad \text{(S88)}$$

where $Q_{th}$ denotes the magnitude of the threshold sheet charge density. Thus, near the onset of accumulation, the terminal threshold voltage $V_{th}$ differs only weakly from the corresponding surface potential reference $\psi_{MoS_2} = \psi_{th}$. In this work, the partition term $Q_{th}/C_{ins}$ is therefore neglected and $\Delta V_{FB} = 0$ is assumed, yielding

$$V_{th} \approx \psi_{th}. \quad \text{(S89)}$$

These additional terms can be readily included in Eq. S86 without altering the scaling trends discussed here. At higher carrier concentrations, however, the partition term $Q/C_{ins}$ increases linearly with $n_{2D}$ and can no longer be neglected, leading to an increasing separation between the terminal and channel voltage scales.

Figure S11A shows that the intrinsic $MoS_2$ channel capacitance $C_{MoS_2}(\psi_{MoS_2})$ increases with surface potential and asymptotically approaches the ideal 2D quantum-capacitance limit $C_Q$ as the channel becomes degenerate, while the inset reports the sheet carrier density $n_{2D}$ on a logarithmic scale. Figure S11B illustrates the physical origin of the centroid penalty. The electron distribution across the $MoS_2$ thickness is nearly symmetric at $\psi_{MoS_2} = \psi_{th} - 0.3$ V and becomes gate-skewed at +0.3 V, indicating a bias-induced charge-centroid displacement toward the gate interface. This centroid shift reduces $\mathrm{d}n_{2D}/\mathrm{d}\psi_{MoS_2}$ and lowers $C_{MoS_2}$ below the ideal quantum-capacitance limit, even in the degenerate regime.

The combined impact of finite DOS and centroid effects is reflected in Fig. S11C, which plots the terminal gate capacitance $C_G$ versus $V_G - V_{th}$ for EOT = 5 Å. At low overdrive, $C_G$ is limited by finite quantum capacitance, whereas at higher overdrive centroid-induced electrostatics further reduce $C_{MoS_2}$ and thus $C_G$ through Eq. S79.

Using Eq. S82, the extracted $MoS_2$ channel CET contribution is shown in Fig. S11D for three configurations. In the absence of a vdW gap ($EOT_{ins}$ = 5 Å), the $MoS_2$ channel CET lies in the range 2.5–3.5 Å at a representative logic overdrive $V_G - V_{th} \approx 0.4$ V. This overdrive is consistent with IRDS targets beyond 2030, where $V_G \approx V_{DD} = 0.6$ V and $V_{th,sat} \approx 0.15$–0.27 V, yielding an effective overdrive $V_{DD} - V_{th,sat} \sim 0.33$–0.45 V depending on performance class.

Introducing a finite vdW gap ($t_{vdW}$ = 1.42 Å, corresponding to $EOT_{vdW} \approx 2.7$ Å) reduces the fraction of gate voltage coupled to the $MoS_2$ surface potential, resulting in a slight deviation from the quantum-capacitance limit and a modest increase in the extracted $CET_{MoS_2}$. A further increase in the vdW gap ($t_{vdW} = 1.5 \times 1.42$ Å), accompanied by a reduced effective permittivity ($1/\kappa_{vdW} = 1 - c/t_{vdW},\ c = 0.72$ Å), yields $EOT_{vdW} \approx 5.4$ Å and produces a more pronounced CET increase.

Even in the atomically thin limit, the $MoS_2$ channel CET contribution remains finite and constitutes an intrinsic electrostatic scaling constraint. In the main manuscript, we therefore adopt $CET_{MoS_2}$ = 3 Å as a representative reference value. Consequently, meeting IRDS CET targets beyond 2030 requires the combined insulator and vdW gap contribution to satisfy $EOT_{ins} + EOT_{vdW} \lesssim$ 6 Å.

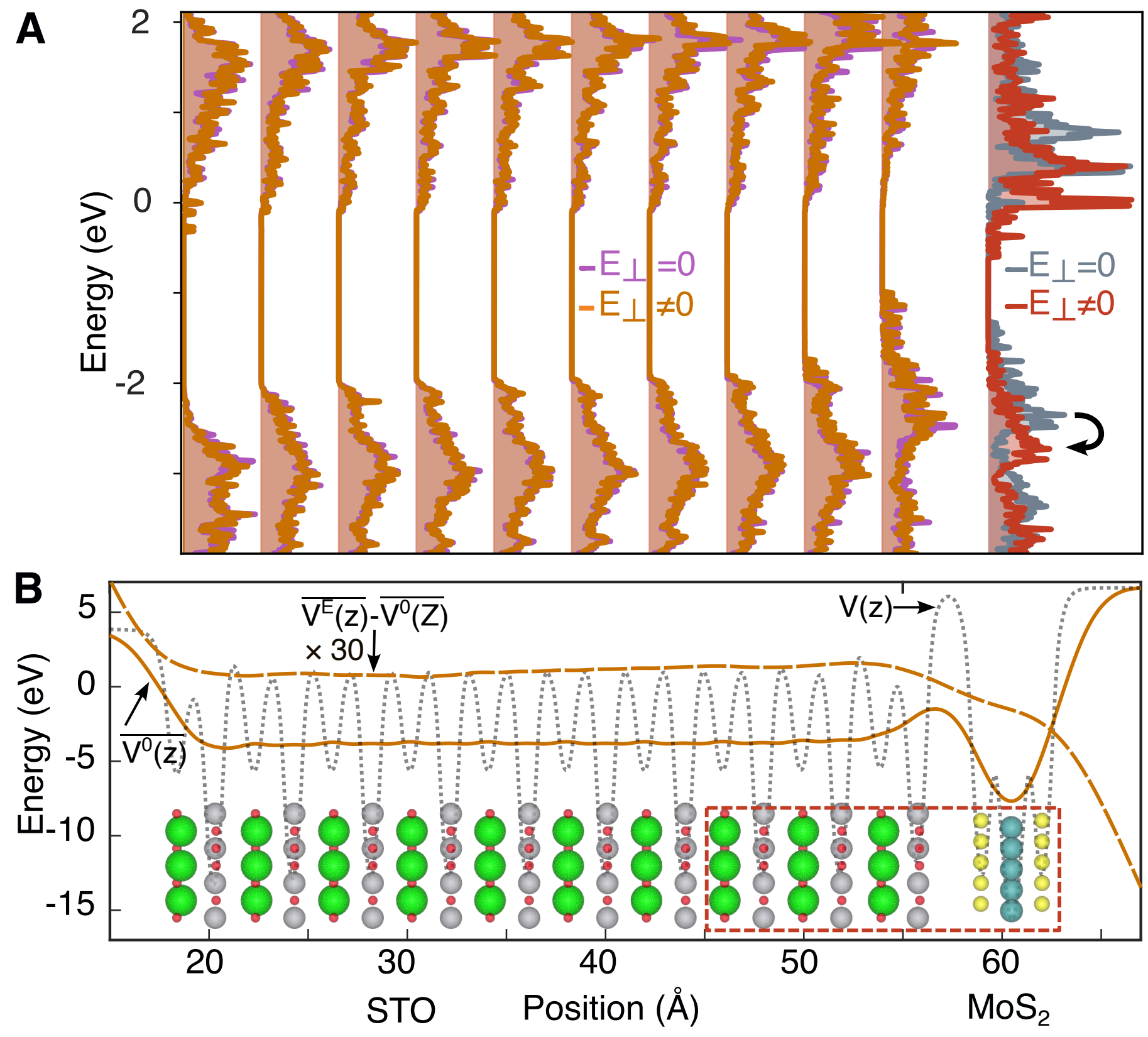

**Figure S1**: **Layer-resolved electronic structure and electrostatics of the STO–$MoS_2$. (A)** Layer-resolved density of states (DOS) showing interfacial hybridization between S atoms in $MoS_2$ and the terminating STO layer. In the absence of an external field, the DOS in the central STO layers remains bulk-like, whereas pronounced interfacial modifications appear under applied field due to reduced dielectric screening at the interface. **(B)** Planar-averaged electrostatic potential $V(z)$, including the atomistic potential, the macroscopic zero-field potential $\overline{V^0(z)}$, and the field-induced potential change $\overline{V^E(z)} - \overline{V^0(z)}$. $TiO_2$ termination is used at the $MoS_2$ interface and SrO termination on the opposite surface to preserve stoichiometry. The flat potential in bulk STO reflects its high permittivity, while sharp variations near the interface and within the vdW gap indicate low-permittivity regions.

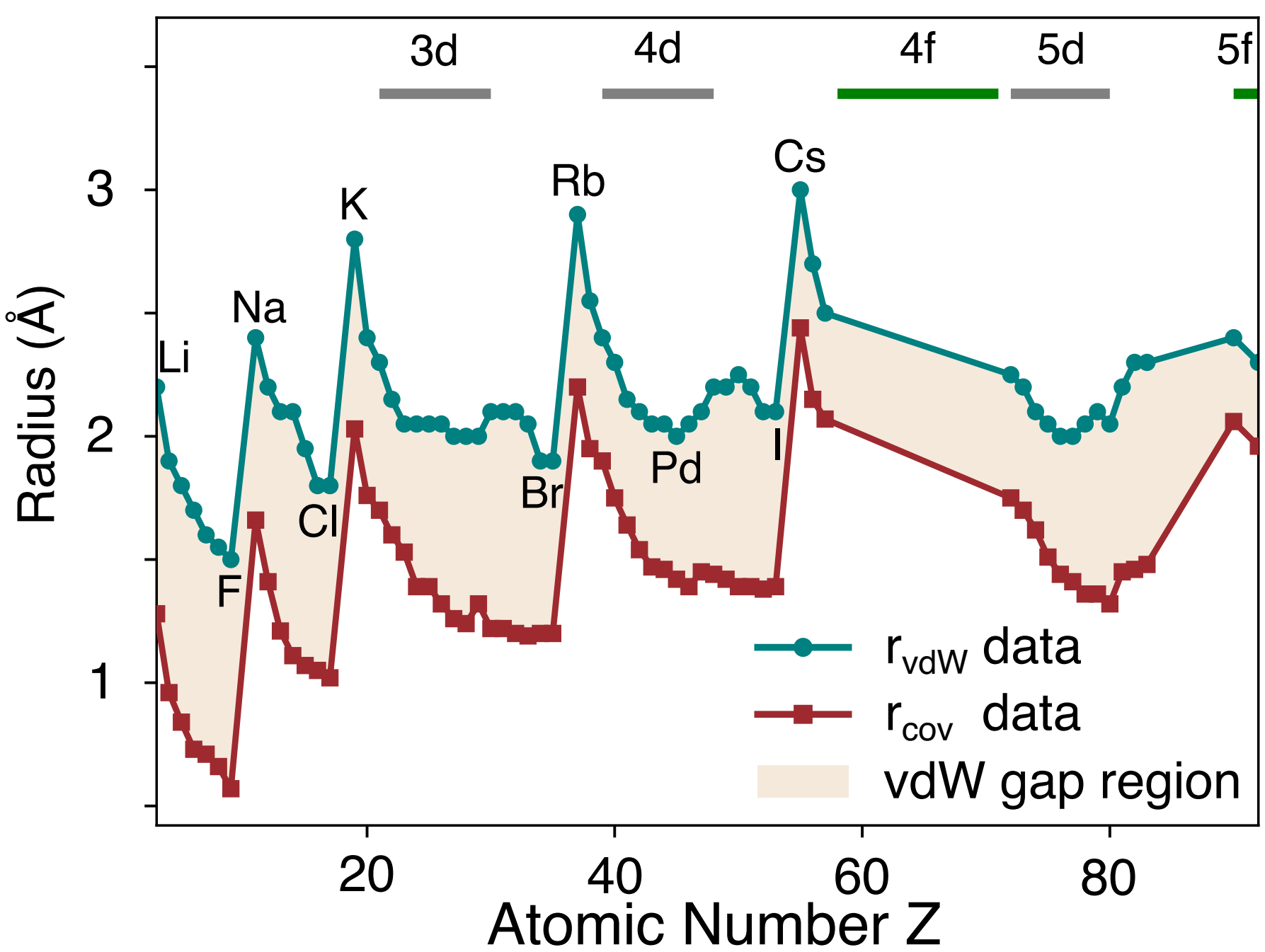

**Figure S2**: **Covalent and van der Waals radii used for vdW-gap estimation.** Covalent radii $r_{\text{cov}}$ and crystal van der Waals radii $r_{\text{vdW}}$ are shown as a function of atomic number $Z$ for the elements listed in Table S1. Their difference defines the single-atom vdW contribution $\Delta_i = r_{\text{vdW}} - r_{\text{cov}}$, which serves as the atomic building block for the binary vdW-gap definition $\Delta_{ij} = \Delta_i + \Delta_j$. Data for $r_{\text{cov}}$ are taken from Cordero *et al.* (*43*) and for $r_{\text{vdW}}$ from Batsanov (*44*). The resulting binary combinations define the chemistry-agnostic vdW-gap thickness distribution shown in Fig. 2D.

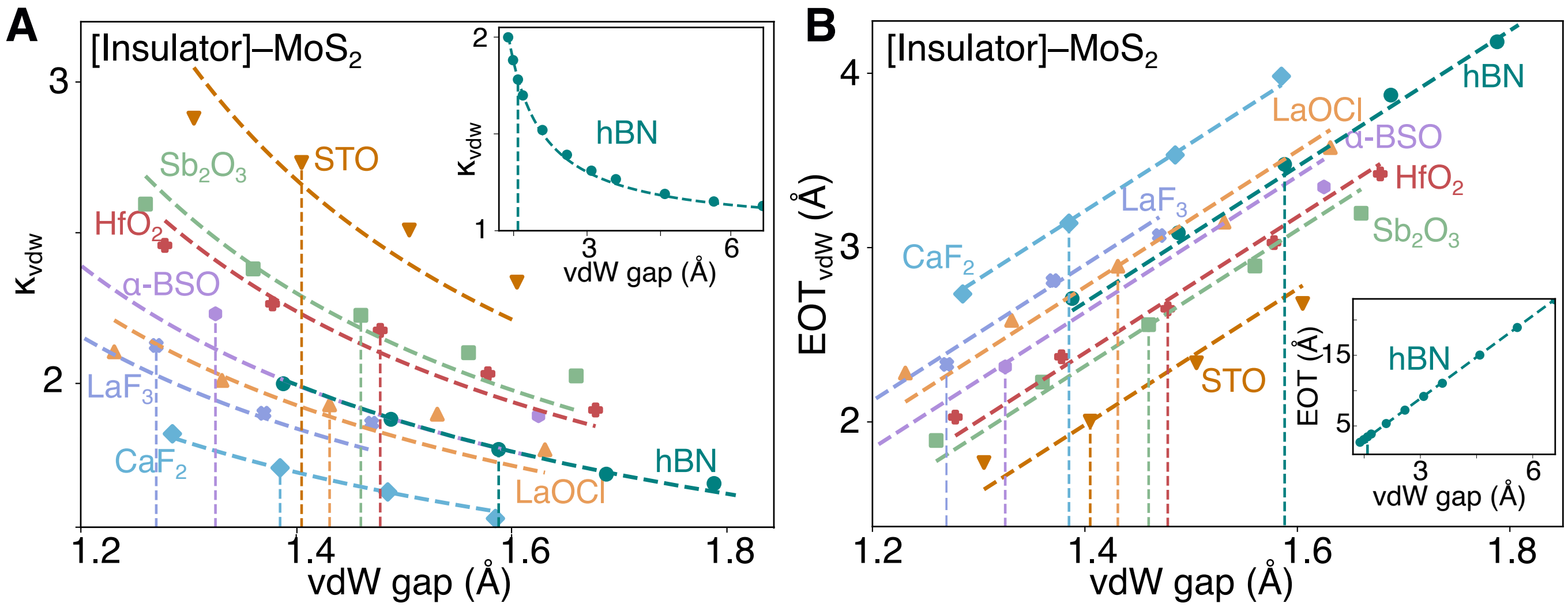

**Figure S3**: **Dependence of vdW gap permittivity and EOT on interfacial separation. (A)** Effective permittivity versus vdW gap thickness for various insulator–MoS$_2$ stacks from ab initio calculations, showing that narrower gaps enhance polarization and increase permittivity. The vdW gap thickness was artificially varied to assess how the gap permittivity changes with separation. The inset highlights results for hBN at increased vdW gap thicknesses. Vertical dashed lines indicate the DFT-calculated vdW gap thickness. **(B)** EOT contribution of the vdW gap for the same systems. The EOT increases nearly linearly with vdW gap thickness; even a 1.4 Å gap adds approximately 2.7 Å to the total EOT.

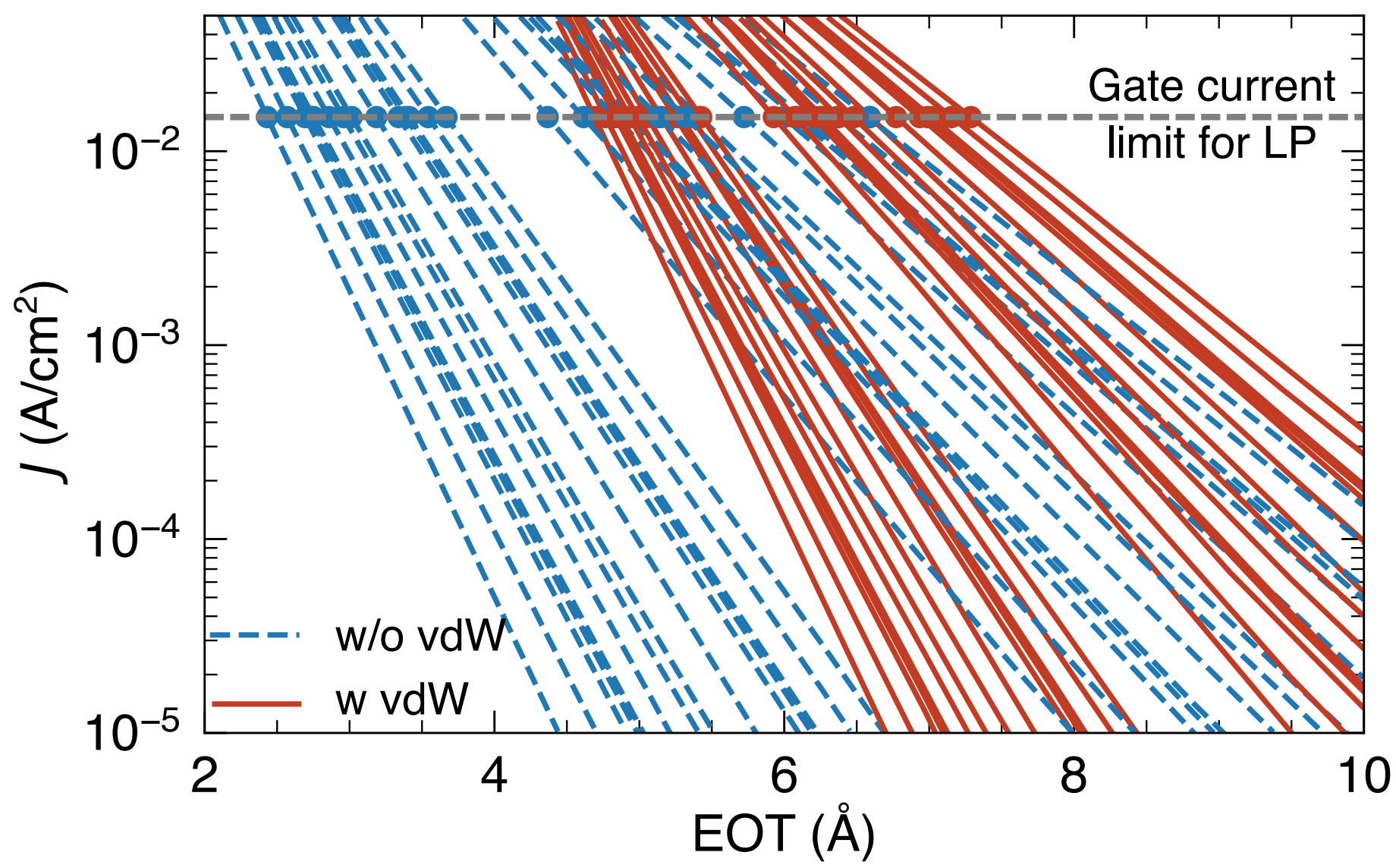

**Figure S4**: **Determination of the minimum achievable EOT.** The tunneling current is computed as a function of EOT for each $MoS_2$–insulator stack using all permutations of the minimum, nominal, and maximum values of the relevant material parameters (permittivity, effective mass, and conduction-band offset), as listed in Table S3. For each parameter set, the minimum achievable EOT is defined as the point where the simulated current intersects the target leakage current. The analysis is performed both without and with a vdW gap. The highlighted markers indicate the minimum EOT extracted for each parameter combination, from which the minimum, median, maximum, and overall range are obtained. Representative results for $CaF_2$–$MoS_2$ are shown.

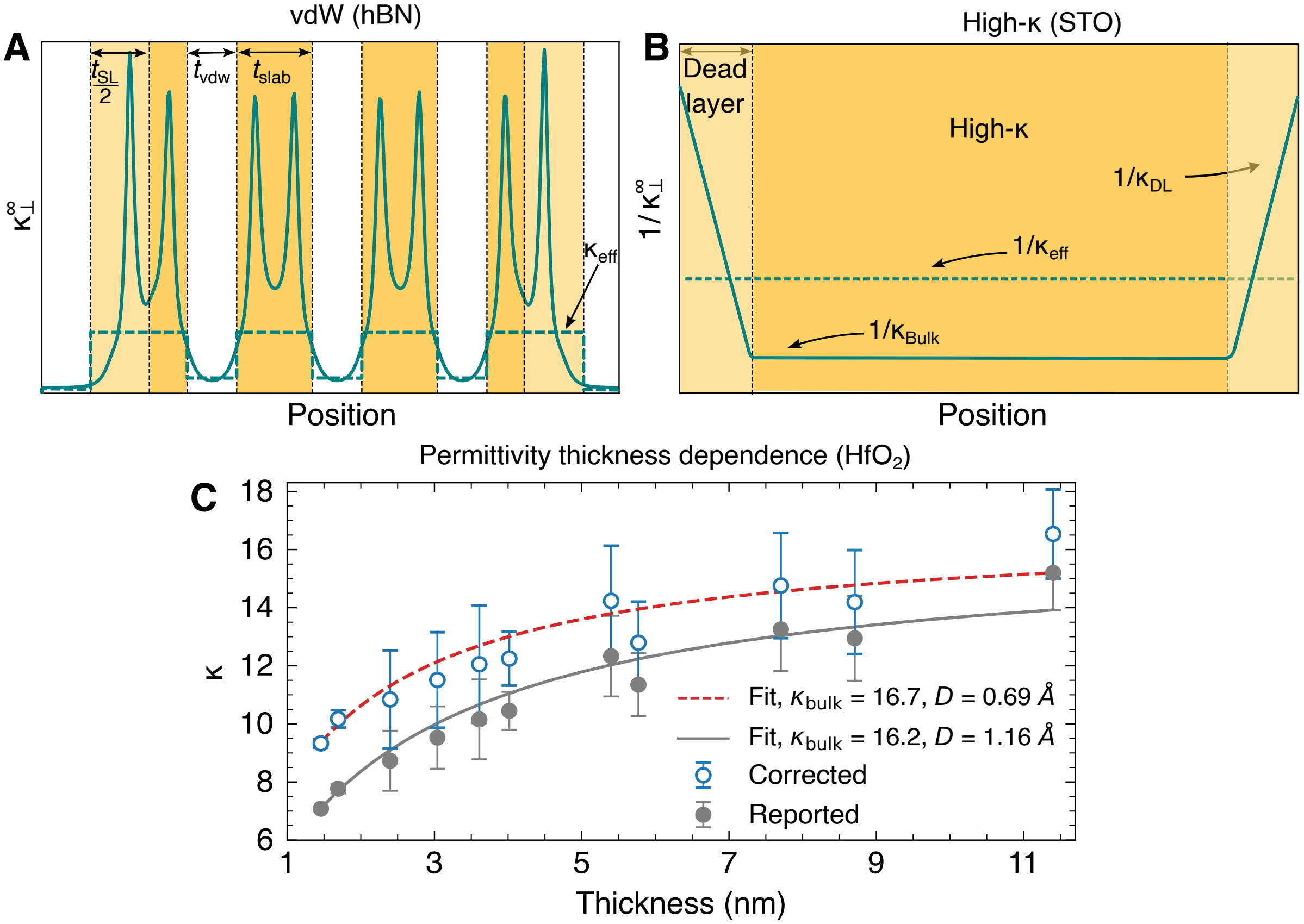

**Figure S5**: **Effective permittivity profiles in van der Waals dielectrics and interfacial dead layers.** (A) Schematic of the effective permittivity profile, $\kappa(z)$, in a layered vdW crystal, showing maxima near atomic planes and minima in the interlayer vdW gaps. The region inside the material, labeled as the slab (filled in orange), represents the atomic layer plus the covalent radii of atoms at the interfaces and is characterized by an effective permittivity $\kappa_{\text{slab}}$. The vdW gap between slabs is shown in white and has an effective permittivity $\kappa_{\text{vdW}}$. Half-layers at the interfaces, which exhibit different polarizability compared to the inner layers, are indicated by softer orange regions. Together, these two half-layers form the single-layer (SL) region, characterized by an effective permittivity $\kappa_{\text{SL}}$. hBN is used as a representative vdW dielectric. (B) Schematic of a high-$\kappa$ insulator with dead layers at both interfaces; STO is used as a representative high-$\kappa$ dielectric. The inverse effective permittivity profile, $1/\kappa(z)$, varies approximately linearly near the interfaces, transitioning from the bulk value to unity (vacuum-like). (C) Thickness-dependent permittivity of $HfO_2$. Experimental data from Ref. (*32*) are shown together with fitted curves using the dead layer model from Eq. S22. Because the measurements were performed in a MIS structure, a possible vdW gap contribution was assumed. Following the approach described in the Supplementary Material, the permittivity was corrected by removing the vdW gap contribution, and a thickness-dependent model was fitted to extract the bulk permittivity and dead layer parameter $D$.

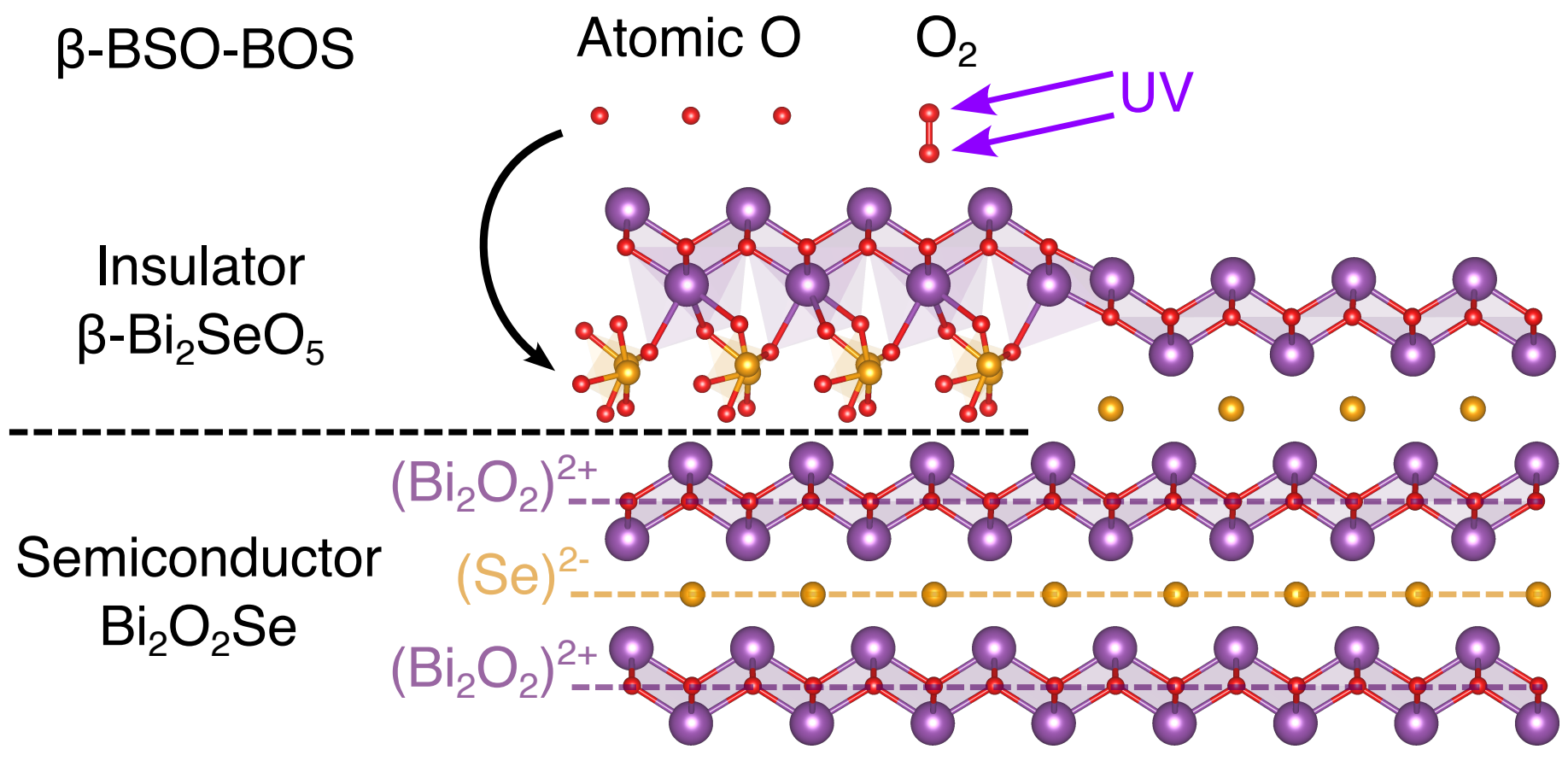

**Figure S6**: **Schematic of a Zipper-like Interface in $\beta$-BSO-BOS Heterostructure.** Schematic illustration of a zipper-like interface formed in the $\beta$-$Bi_2SeO_5$–$Bi_2O_2Se$ ($\beta$-BSO–BOS) heterostructure. The interface creates intermediate bonds that are stronger than vdW interactions but not fully covalent, effectively avoiding interfacial gaps. $\beta$-BSO is the native oxide of BOS.

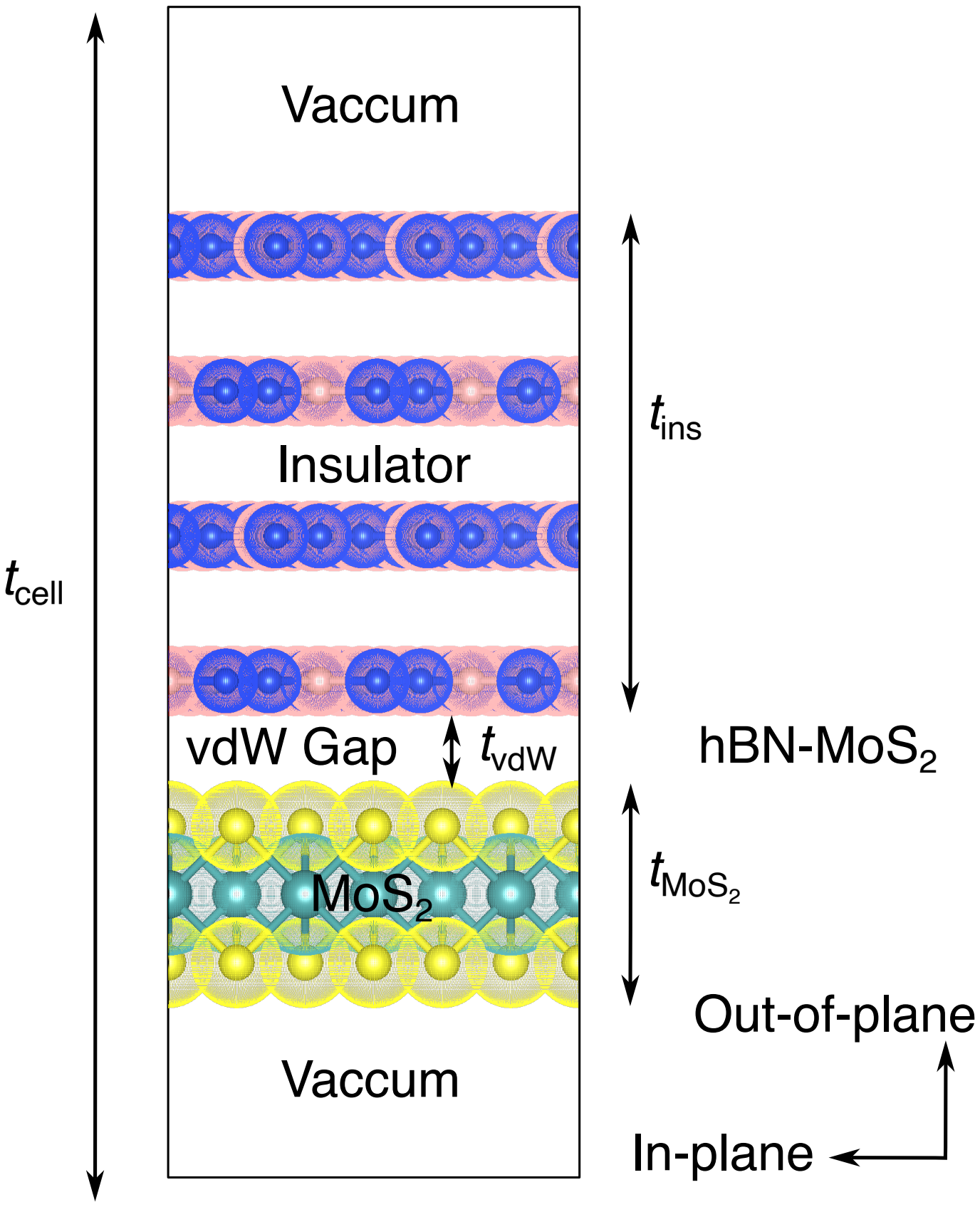

**Figure S7**: **Simulation Cell Geometry for Macroscopic Dielectric Calculations**. Schematic diagram of the simulation supercell consisting of an insulator, vdW gap, and $MoS_2$ layer, each surrounded by vacuum. The total length of the supercell is $L_{\text{cell}}$, with labeled thicknesses $L_{\text{ins}}$, $L_{\text{vdW}}$, and $L_{\text{MoS}_2}$, indicating the lengths of the insulator, vdW gap, and $MoS_2$, respectively. Layer thicknesses are illustrative and not drawn to scale. This layered geometry is used in the microscopic and macroscopic dielectric analysis described in this and the following section.

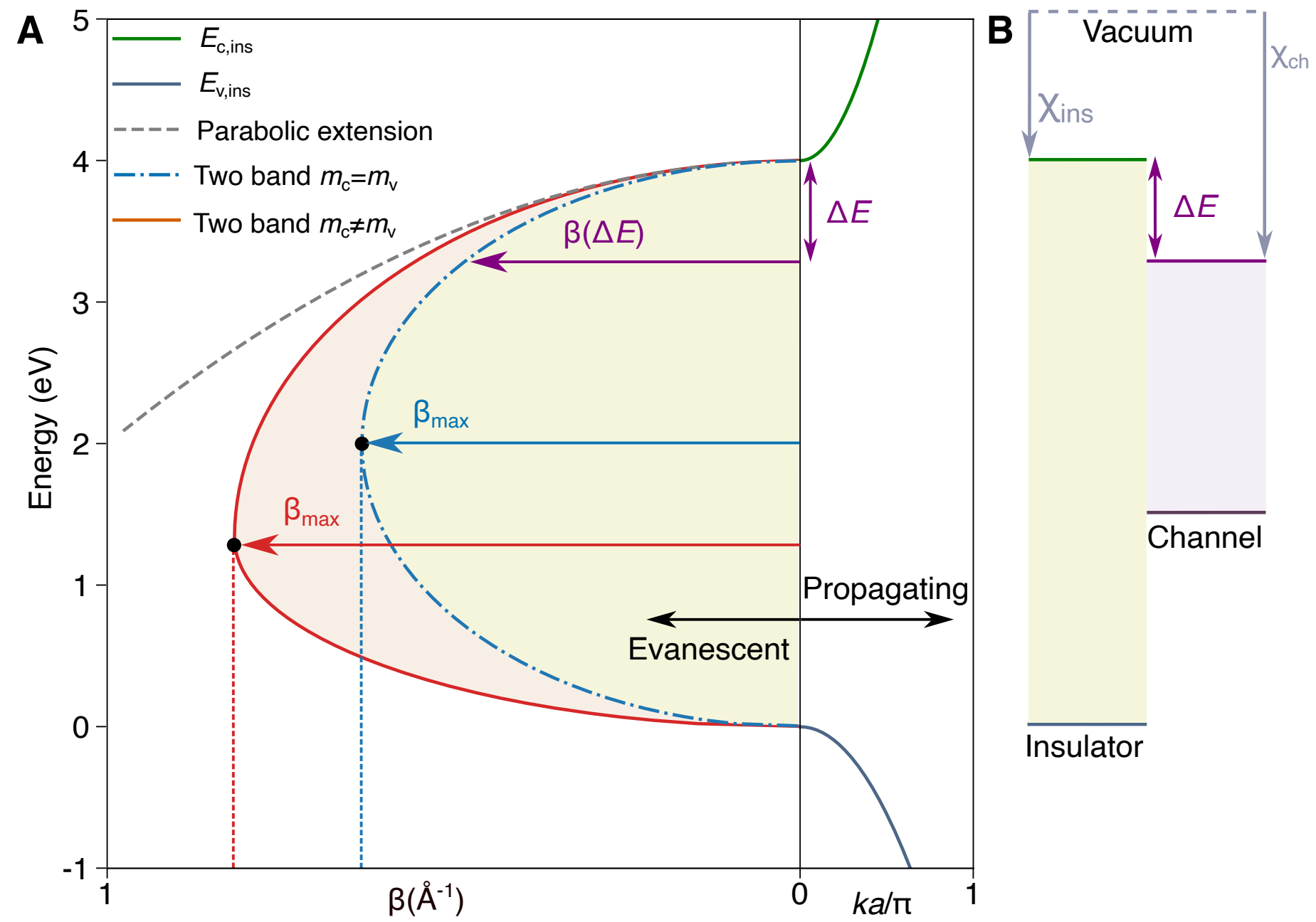

**Figure S8**: **Complex Band Structure Models and Tunneling Decay Parameters**. (A) Schematic comparison of three complex band-structure models. Right: real $k$-space dispersion for the conduction and valence bands. Left: corresponding inverse decay length $\beta(E)$. Dashed curve: single-parabolic continuation. Dash-dotted curve: symmetric two-band (Franz) model. Solid curve: asymmetric two-band model illustrating the branch-point shift that arises from unequal electron and hole effective masses. The purple curve $\beta(\Delta E)$ marks the inverse decay length evaluated at an energy offset $\Delta E$ measured from the conduction-band minimum. (B) Schematic band alignment at an insulator/semiconductor interface. The electron affinities, $\chi$, of the two materials determine the conduction-band offset, $\Delta E$, which sets the height of the tunneling barrier that electrons must overcome to traverse the insulator and reach the channel's conduction-band edge. The inverse decay length associated with this barrier height is shown as the purple curve in part (A).

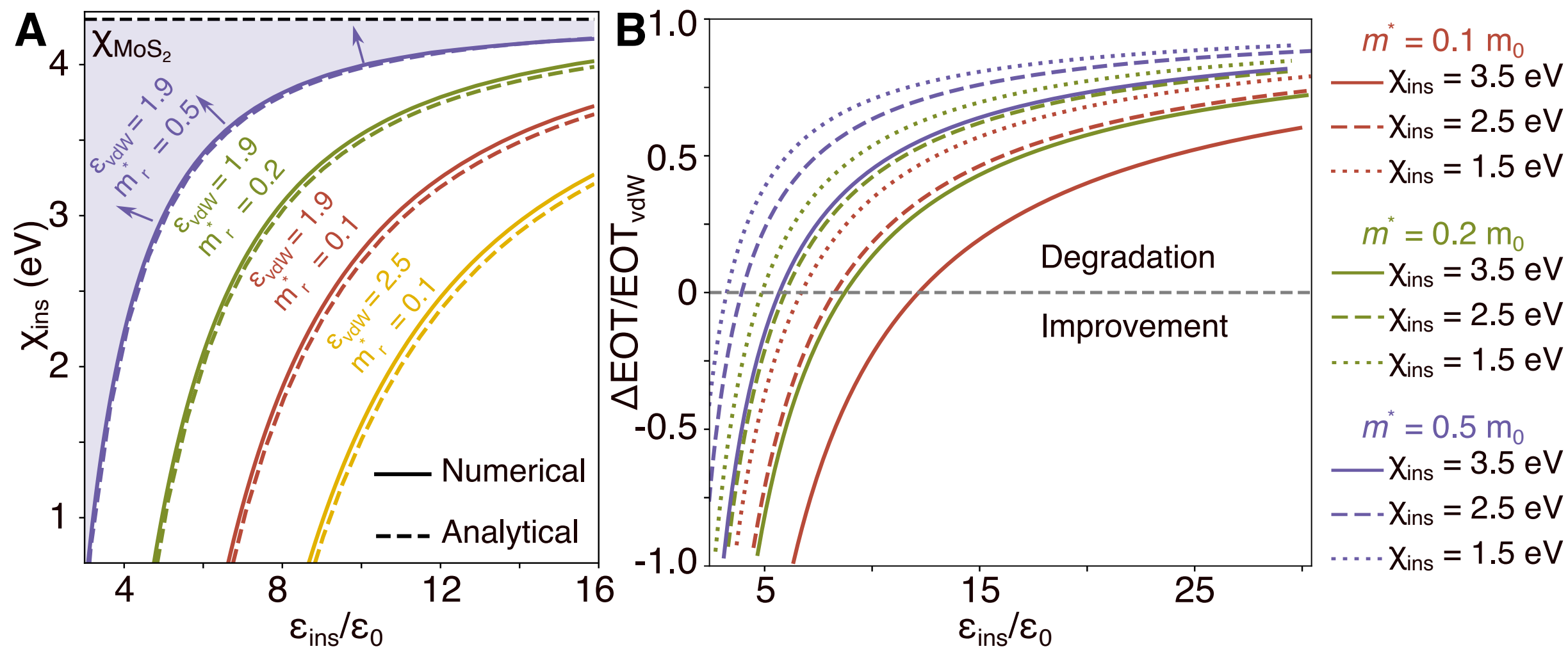

**Figure S9**: **Analytical Boundaries and EOT Impact of vdW Gaps in Insulator Scaling**. (A) Analytical boundaries defined by Eq. S67 for different values of $m_r^*$. Points in the region above each curve correspond to combinations of $(\kappa_{\text{ins}}, \chi_{\text{ins}})$ where introducing a vdW gap reduces the minimum achievable EOT. Dashed curves: analytical model; solid curves: full numerical results. (B) $\Delta\text{EOT}/\text{EOT}_{\text{vdW}}$ vs. insulator permittivity for various $\chi_{\text{ins}}$ and $m^*_{\text{ins}}$. Negative values indicate improved EOT performance. $\kappa_{\text{vdW}} = 2$, $\alpha = 0.8$, and $\chi_{\text{MoS}_2} = 4.3\,\text{eV}$.

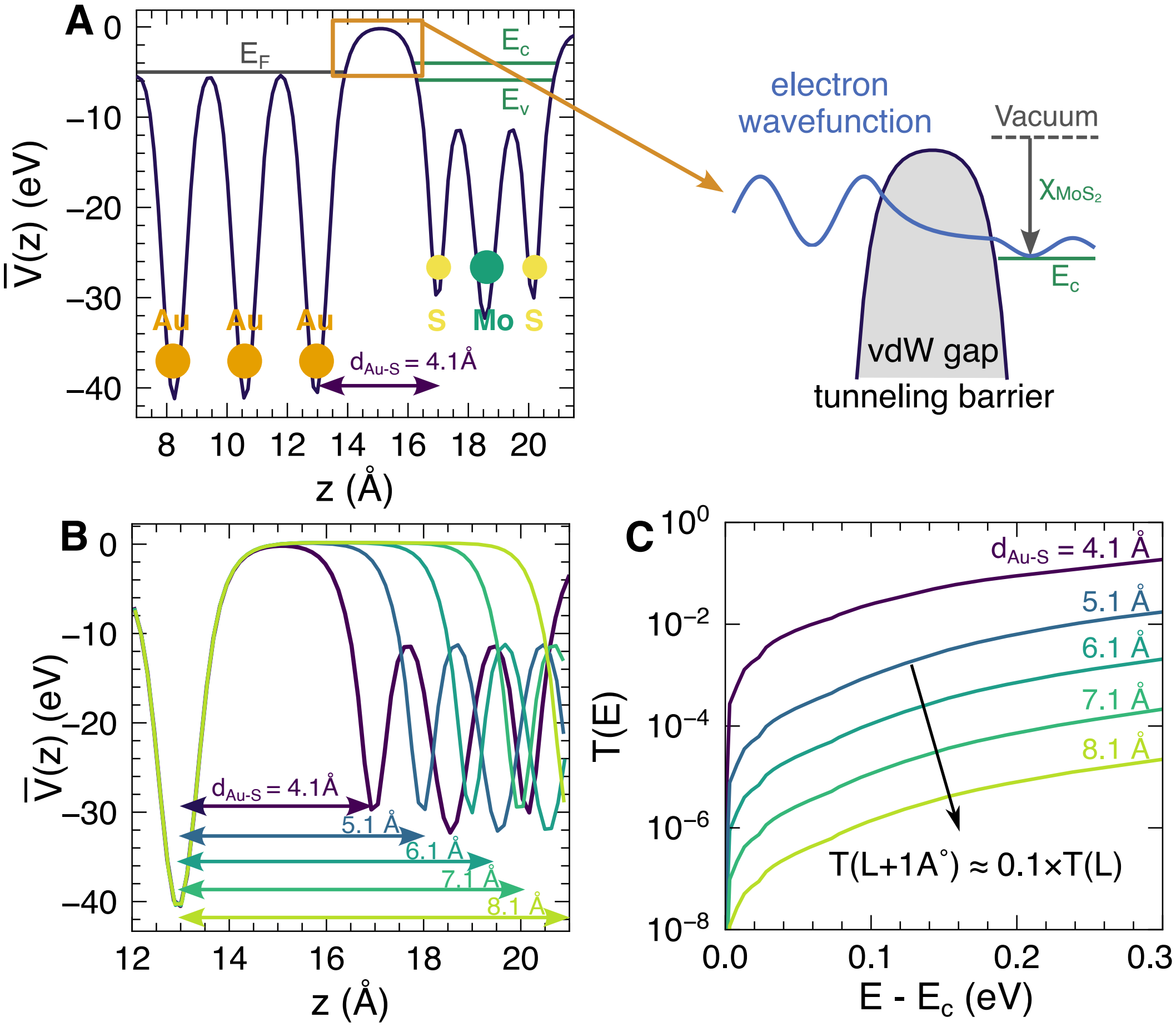

**Figure S10**: **Tunneling across the vdW gap at an Au–$MoS_2$ contact.** **(A)** Planar-averaged electrostatic potential $\overline{V}(z)$ across the interface. Since the vdW gap maximum approaches the vacuum level, the electron-injection barrier into $MoS_2$ is approximately its electron affinity. **(B)** Zoom of $\overline{V}(z)$ in the vdW region for gaps increased from 4.1 to 8.1 Å in 1 Å steps. **(C)** Tunneling probability at the $MoS_2$ conduction-band edge versus vdW gap thickness, showing an approximately tenfold reduction for each additional 1 Å of separation.

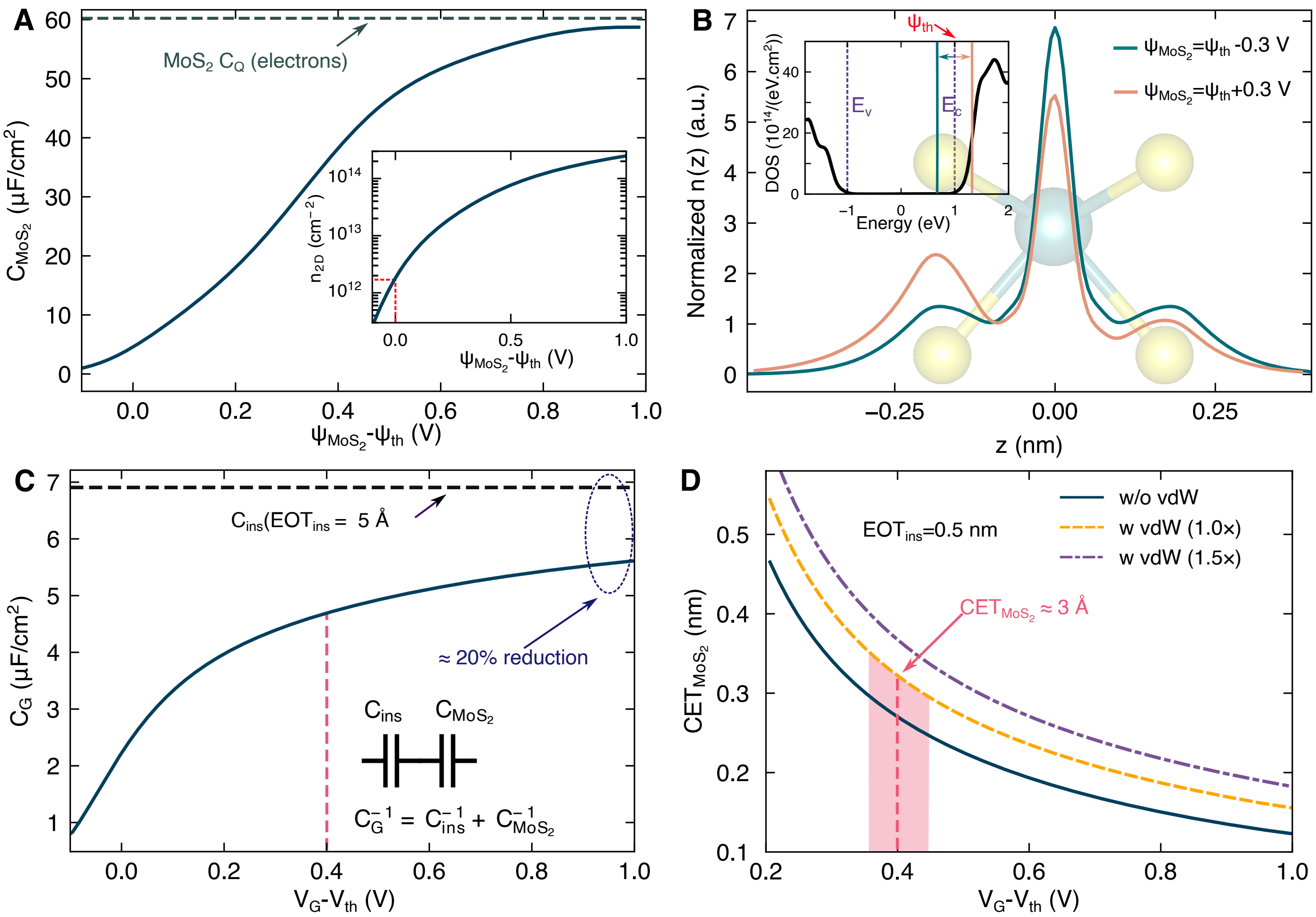

**Figure S11**: **$MoS_2$ channel capacitance and CET extraction used for EOT scaling analysis.** **(A)** Intrinsic $MoS_2$ channel capacitance $C_{\mathrm{MoS_2}} = q\,\mathrm{d}n_{2\mathrm{D}}/\mathrm{d}\psi_{\mathrm{MoS_2}}$ versus $\psi_{\mathrm{MoS_2}} - \psi_{\mathrm{th}}$, together with the 2D quantum-capacitance limit from Eq. S84. The inset shows the sheet carrier density $n_{2\mathrm{D}}$ on a logarithmic scale; guide lines at $\psi_{\mathrm{MoS_2}} = \psi_{\mathrm{th}}$ indicate $n_{2\mathrm{D}} \approx 2 \times 10^{12}\ \mathrm{cm}^{-2}$, used as a practical threshold-like reference for the onset of accumulation. **(B)** Normalized electron distribution across monolayer $MoS_2$ at $\psi_{\mathrm{MoS_2}} = \psi_{\mathrm{th}} \pm 0.3$ V, illustrating bias-induced charge-centroid displacement toward the gate interface. The threshold reference is defined as $\psi_{\mathrm{th}} = E_G/(2q)$ (midgap energy reference); at $\psi_{\mathrm{MoS_2}} = \psi_{\mathrm{th}}$, the Fermi level aligns with the conduction-band edge $E_{\mathrm{c}}$, as indicated in the inset. **(C)** Terminal gate capacitance $C_{\mathrm{G}}$ versus $V_{\mathrm{G}} - V_{\mathrm{th}}$ for $\mathrm{EOT_{ins}} = 5$ Å in the absence of a vdW gap. **(D)** Extracted $MoS_2$ channel CET contribution for three electrostatic configurations: (i) no vdW gap ($\mathrm{EOT_{ins}} = 5$ Å), (ii) a vdW gap of $t_{\mathrm{vdW}} = 1.42$ Å ($\mathrm{EOT_{vdW}} \approx 2.7$ Å), and (iii) an enlarged vdW gap $t_{\mathrm{vdW}} = 1.5 \times 1.42$ Å, yielding $\mathrm{EOT_{vdW}} \approx 5.4$ Å. A representative value of $\mathrm{CET_{MoS_2}} = 3$ Å is used in the main manuscript to evaluate ultimate insulator and vdW gap EOT scaling limits. In **(C)** and **(D)**, a vertical guide line marks a representative logic overdrive $V_{\mathrm{G}} - V_{\mathrm{th}} = 0.4$ V, consistent with IRDS projected core-logic targets beyond 2030.

**Table S1**: **Tabulated atomic radii used for vdW-gap estimation.** The table lists covalent radii $r_{\mathrm{cov}}$ (*43*) and crystal van der Waals radii $r_{\mathrm{vdW}}$ (*44*), together with the derived single-atom vdW contribution $\Delta_i = r_{\mathrm{vdW}} - r_{\mathrm{cov}}$. These atomic quantities form the basis for the binary vdW-gap definition $\Delta_{ij} = \Delta_i + \Delta_j$, evaluated over unique elemental pairs. Statistical analysis over the atomic set yields a characteristic vdW-gap scale that is independent of specific bonding chemistry and underlies the vdW-gap thickness distribution presented in Fig. 2D. All lengths are in Å.

| Elem. | $Z$ | $r_{\mathrm{cov}}$ | $r_{\mathrm{vdW}}$ | $\Delta_i$ | Elem. | $Z$ | $r_{\mathrm{cov}}$ | $r_{\mathrm{vdW}}$ | $\Delta_i$ | Elem. | $Z$ | $r_{\mathrm{cov}}$ | $r_{\mathrm{vdW}}$ | $\Delta_i$ |
|---|---|---|---|---|---|---|---|---|---|---|---|---|---|---|
| Li | 3 | 1.28 | 2.20 | 0.92 | Mn | 25 | 1.39 | 2.05 | 0.66 | Ag | 47 | 1.45 | 2.10 | 0.65 |
| Be | 4 | 0.96 | 1.90 | 0.94 | Fe | 26 | 1.32 | 2.05 | 0.73 | Cd | 48 | 1.44 | 2.20 | 0.76 |
| B | 5 | 0.84 | 1.80 | 0.96 | Co | 27 | 1.26 | 2.00 | 0.74 | In | 49 | 1.42 | 2.20 | 0.78 |
| C | 6 | 0.73 | 1.70 | 0.97 | Ni | 28 | 1.24 | 2.00 | 0.76 | Sn | 50 | 1.39 | 2.25 | 0.86 |
| N | 7 | 0.71 | 1.60 | 0.89 | Cu | 29 | 1.32 | 2.00 | 0.68 | Sb | 51 | 1.39 | 2.20 | 0.81 |
| O | 8 | 0.66 | 1.55 | 0.89 | Zn | 30 | 1.22 | 2.10 | 0.88 | Te | 52 | 1.38 | 2.10 | 0.72 |
| F | 9 | 0.57 | 1.50 | 0.93 | Ga | 31 | 1.22 | 2.10 | 0.88 | I | 53 | 1.39 | 2.15 | 0.76 |
| Ne | 10 | 0.58 | 1.54 | 0.96 | Ge | 32 | 1.20 | 2.10 | 0.90 | Xe | 54 | 1.40 | 2.16 | 0.76 |
| Na | 11 | 1.66 | 2.40 | 0.74 | As | 33 | 1.19 | 2.05 | 0.86 | Cs | 55 | 2.44 | 3.20 | 0.76 |
| Mg | 12 | 1.41 | 2.20 | 0.79 | Se | 34 | 1.20 | 1.90 | 0.70 | Ba | 56 | 2.15 | 2.80 | 0.65 |
| Al | 13 | 1.21 | 2.10 | 0.89 | Br | 35 | 1.20 | 1.90 | 0.70 | La | 57 | 2.07 | 2.70 | 0.63 |
| Si | 14 | 1.11 | 2.10 | 0.99 | Kr | 36 | 1.16 | 2.02 | 0.86 | Hf | 72 | 1.78 | 2.25 | 0.47 |
| P | 15 | 1.07 | 1.95 | 0.88 | Rb | 37 | 2.20 | 3.00 | 0.80 | Ta | 73 | 1.64 | 2.20 | 0.56 |
| S | 16 | 1.05 | 1.80 | 0.75 | Sr | 38 | 1.95 | 2.60 | 0.65 | W | 74 | 1.62 | 2.10 | 0.48 |
| Cl | 17 | 1.02 | 1.80 | 0.78 | Y | 39 | 1.90 | 2.50 | 0.60 | Ir | 77 | 1.41 | 2.00 | 0.59 |
| Ar | 18 | 1.06 | 1.88 | 0.82 | Zr | 40 | 1.75 | 2.30 | 0.55 | Pt | 78 | 1.36 | 2.05 | 0.69 |
| K | 19 | 2.03 | 2.80 | 0.77 | Nb | 41 | 1.64 | 2.15 | 0.51 | Au | 79 | 1.36 | 2.05 | 0.69 |
| Ca | 20 | 1.76 | 2.40 | 0.64 | Mo | 42 | 1.54 | 2.10 | 0.56 | Hg | 80 | 1.32 | 2.05 | 0.73 |
| Sc | 21 | 1.70 | 2.30 | 0.60 | Tc | 43 | 1.47 | 2.05 | 0.58 | Tl | 81 | 1.45 | 2.20 | 0.75 |
| Ti | 22 | 1.60 | 2.15 | 0.55 | Ru | 44 | 1.46 | 2.05 | 0.59 | Pb | 82 | 1.46 | 2.30 | 0.84 |
| V | 23 | 1.53 | 2.05 | 0.52 | Rh | 45 | 1.42 | 2.05 | 0.63 | Bi | 83 | 1.48 | 2.30 | 0.82 |
| Cr | 24 | 1.39 | 2.05 | 0.66 | Pd | 46 | 1.39 | 2.05 | 0.66 | Th | 90 | 2.06 | 2.40 | 0.34 |

**Table S2**: **vdW gap geometry, dielectric response, and energetics in insulator–$MoS_2$ stacks.** All quantities reported in this table are extracted for insulator–$MoS_2$ interfaces. The equilibrium atomic interface separation $d_{int}$ is obtained from relaxed ab initio interface structures and corresponds to the distance between the two facing atomic planes across the interface. The effective vdW gap thickness $t_{vdW}$ and permittivity $\kappa_{vdW}$ are obtained from DFT calculations of the interfacial vdW region. The parameter $c$ is obtained from fitting the thickness-dependent permittivity model $1/\kappa_{vdW} = 1 - c/t_{vdW}$. $E_{bind}$ denotes the interfacial binding energy density. Average values across all materials are $\overline{d}_{int}$ = 2.95 Å, $\overline{t}_{vdW}$ = 1.42 Å, $\overline{\kappa}_{vdW}$ = 2.06, $\overline{c}$ = 0.72 Å, $\overline{\text{EOT}}$ = 2.73 Å, and $\overline{E}_{bind}$ = 23.29 meV/Å$^2$.

| **Insulator** | $LaF_3$ | LaOCl | $Sb_2O_3$ | $CaF_2$ | $\alpha$-BSO | $HfO_2$ | STO | hBN | **Average (Data)** |
|---|---|---|---|---|---|---|---|---|---|
| $d_{int}$ (Å) | 2.70 | 3.11 | 2.97 | 2.87 | 2.67 | 2.94 | 3.00 | 3.34 | 2.95 |
| $t_{vdW}$ (Å) | 1.27 | 1.43 | 1.46 | 1.39 | 1.33 | 1.48 | 1.41 | 1.59 | 1.42 |
| $\kappa_{vdW}$ | 2.03 | 1.89 | 2.18 | 1.72 | 2.11 | 2.10 | 2.66 | 1.79 | 2.06 |
| $c$ (Å) | 0.64 | 0.67 | 0.79 | 0.58 | 0.70 | 0.77 | 0.88 | 0.70 | 0.72 |
| EOT (Å) | 2.44 | 2.95 | 2.61 | 3.14 | 2.45 | 2.74 | 2.06 | 3.47 | 2.73 |
| $E_{bind}$ (meV/Å$^2$) | 28.55 | 18.73 | 19.85 | 18.64 | 19.44 | 25.55 | 33.00 | 22.57 | 23.29 |

**Table S3**: Material parameters used for tunneling calculations. Listed are the bulk relative permittivity $\kappa$, electron affinity $\chi$, tunneling effective mass $m^*/m_0$, representative band gap $E_g$, and dead-layer thickness $D$, where applicable. A thickness-dependent permittivity (dead-layer) model was applied only to STO and $HfO_2$. For STO, $D$ was taken directly from Ref. (*18*). For $HfO_2$, thickness-dependent capacitance data from Ref. (*32*) were corrected for the vdW gap contribution (Fig. S5C).

| Material | $\kappa$ | $\chi$ (eV) | $m^*/m_0$ | $E_g$ (eV) | $D$ (Å) |
|---|---|---|---|---|---|
| $\beta$-BSO ($Bi_2Se_2O_5$) | 14–16 | 2.0–2.4 | 0.6–0.8 | 3.6–3.7 | – |
| $LaF_3$ | 14–16.5 | 2.3 | 0.90 | 9.7 | – |
| LaOCl | 10.8–13.8 | 2.3 | 1.12 | 5.5 | – |
| $Sb_2O_3$ | 11.5 | 2.9–3.2 | 0.45–1.04 | 4.4 | – |
| $\alpha$-BSO ($Bi_2Se_2O_5$) | 10.8–16.2 | 1.89–3.0 | 0.50–1.65 | 3.5-3.9 | – |
| STO | 270 | 3.3–4.1 | 0.10–0.50 | 3.3 | 1.6 |
| $CaF_2$ | 6.8–8.4 | −0.15–2.9 | 0.31–1.20 | 11.8 | – |
| $HfO_2$ | 16.7 | 1.75–2.25 | 0.08–0.14 | 5.3 | 0.69 |
| hBN | 3.4–5.1 | 2.10–2.26 | 0.50 | 5.9 | – |